\documentclass[12pt,preprint]{aastex}

\shorttitle{Solar Wind Electron Heating}
\shortauthors{Laming}

\begin{document}

\title{On Collisionless Electron-Ion Temperature
Equilibration in the Fast Solar Wind}


\author{J. Martin Laming\altaffilmark{1}}


\altaffiltext{1}{E. O Hulburt Center for Space Research, Naval Research Laboratory,
Code 7674L, Washington DC 20375 \email{jlaming@ssd5.nrl.navy.mil}}

\begin{abstract}
We explore a mechanism, entirely new to the fast solar wind, of electron
heating by lower hybrid waves to explain the shift to higher charge
states observed in various elements in the fast wind at 1 A.U. relative to
the original coronal hole plasma. This process is a variation on that
previously discussed for two temperature accretion flows by
Begelman \& Chiueh. Lower hybrid waves are generated by gyrating
minor ions (mainly $\alpha$-particles) and become significant once
strong ion cyclotron heating sets in beyond 1.5$R_{\sun}$. In this way
the model avoids conflict with SUMER electron temperature diagnostic
measurements between 1 and 1.5$R_{\sun}$. The principal requirement for
such a process to work is the existence of density gradients in the fast
solar wind, with scale length of similar order to the proton inertial length.
Similar size structures have previously been inferred by other authors
from radio scintillation
observations and considerations of ion cyclotron wave generation by global
resonant MHD waves.
\end{abstract}



\section{Introduction}
The 1990 October launch of Ulysses and the subsequent polar passes in late
1994, 1995, 2000 and 2001 have provided a wealth of new data and insights
concerning solar polar coronal holes and the fast solar wind emanating
from them. The south polar pass of 1994 (during solar minimum)
highlighted a particular problem
concerning the charge states observed in various elements in the fast wind,
which are generally characteristic of higher temperatures than observed
spectroscopically in coronal holes. Coronal holes show very little emission
in spectral lines corresponding to plasma temperatures of $10^6$ K or
greater \citep[e.g.][]{doschek77,david98,wilhelm98}, whereas \citet{geiss95} find
freeze-in temperatures from the O, Si and Fe charge states in the range
$1.2 - 1.5 \times 10^6$ K. Fe in the fast solar wind is dominated by charge
states Fe$^{10+}$ and Fe$^{11+}$, whereas lines from Fe X, XI, and XII are generally not
detectable in spectroscopy of coronal holes; an old result going back to
\citet{doschek77}. \citet{geiss95}
and \citet{ko97} recognized that the electron temperature must increase
outwards as the fast wind flows out of the coronal hole, and were able to
derive a empirical temperature profile, but did not specify a mechanism to
produce such an effect.

The heuristic temperature profile of \citet{ko97} predicts a maximum
in the electron temperature of about $1.5 \times 10^6$ K relatively close to
the solar surface, at about $1.5 R_{\sun}$ heliocentric distance. However
observations in a temperature diagnostic line ratio in Mg~IX \citep{wilhelm98}
out to $1.6 R_{\sun}$ with the SUMER instrument on SOHO failed to detect such a
temperature increase. Consequently, in a series of papers \citep{esser00,esser01,chen03}, the effects of a halo electron
distribution (i.e. a power law tail on the core Maxwellian) that remains consistent
with SUMER electron temperature measurements and differential
flows between ions of the same element were investigated. While carefully chosen
differential flows can explain most of the in situ charge state observations without
the need for extra electron heating, the required speed differentials are so large as
to be implausible, leaving electron heating as the most likely explanation.
The fast solar wind is often observed (at 1 AU)
with $T_{e||}>T_{e\perp}$
\citep{feldman75} and with $T_{e||}/T_{e\perp} > T_{p||}/T_{p\perp}$, and this
electron temperature anisotropy can be described in term of a core Maxwellian
distribution and a halo higher temperature Maxwellian or power law distribution, which
contains typically 5\% of the electrons \citep{marsch91}. It has been suggested that
this halo distribution persists all the way back to the sun \citep{esser00}, and this
might be appealing since the electron temperature diagnostic used by \citet{wilhelm98}
(Mg IX) belongs to an isoelectronic sequence that is in general
insensitive to such a population \citep{keenan84a}, whereas the ionization
balance is not. However the O VI diagnostic used by \citet{david98} should be
sensitive to suprathermal electrons but shows no evidence of their existence
out to 1.3 $R_{\sun}$. In any case,
it seems that such an explanation must fail, because the halo distribution will
transfer energy to the core Maxwellian by Coulomb collisions on a timescale
$\sim\left(10^7-10^8\right)\times T_{e6}^{3/2}/n_e$ s, ($T_{e6}$ is the electron
temperature in $10^6$K, and $n_e$ is the electron density in cm$^{-3}$)
and even with the fastest inferred outflow speeds, the solar wind
would not get very far before a discernable increase in the core Maxwellian
electron temperature should become apparent.

\section{Wave Acceleration and Heating in the Fast Solar Wind}
The acceleration of the fast solar wind is widely believed to occur through
ion cyclotron waves, though the manner in which they are excited remains controversial.
The original concept, that low frequency waves alone
are sufficient to accelerate the fast solar wind to its eventual speed of 700-800
km s$^{-1}$, appears to be adequate for acceleration of protons but not for the minor
ions \citep{ofman01}, which in this case would be accelerated mainly by the
Coulomb drag of the protons. Hence minor ions should flow out of the coronal
hole with lower speeds than the protons.
Renewed indications that ion cyclotron waves are required come from the UVCS
Doppler dimming measurements of the O$^{5+}$ outflow velocity, which are {\em large}
compared to hydrogen, $\sim 300-400$ km s$^{-1}$ at 2.5-3.0 $R_{\sun}$ heliocentric
distance. By contrast a purely low frequency wave driven outflow would produce O$^{5+}$
outflow velocities of less than 100 km s$^{-1}$ at this distance.
Similar indications come from the observation in the fast solar wind of
faster flow speeds for $\alpha$ particles (and in fact all minor ions)
than for protons with Ulysses and Helios
\citep{reisenfeld01,vonsteiger00,neugebauer96}. The Ulysses
and Helios observations only go in to the sun as far as $\sim 1.4R_{\sun}$ heliocentric
distance, but the inference that $\alpha$ particles flow out from the sun at least as
fast as protons in the fast solar wind may be extended to lower altitudes by
spectroscopic observations with SUMER of the He abundance \citep{laming03a}. Here He
abundances at or below the usual fast solar wind value of 5\% relative to H are found,
implying flow speeds at these lower altitudes
at or above the H flow speed to keep the flux of $\alpha$ particles constant.
This is also consistent with recent modelling by \citet{li03}, where by heating
primarily the $\alpha$ particles, a steep transition region, hot corona and
fast wind can be created by high frequency Alfv\'en waves propagating up from
the coronal base. On the other hand \citet{cranmer99a} and \citet{cranmer00}
argue that the high frequency (or ion cyclotron) waves that accelerate the
fast wind must be generated throughout the extended corona, due to their
rapid dissipation once generated, and the observation that ions continue to be
heated in the range $1.5 - 5 R_{\sun}$ heliocentric distance.

We propose an entirely new \citep[to the fast wind; though see][for related
considerations]{schwartz81} means of heating/accelerating electrons.
At a heliocentric distance between 1.5 and 2 $R_{\sun}$, the perpendicular {\em
velocities} of the minor ions start to exceed those of the protons \citep[e.g.][]
{cranmer99}. There are also indications
from observations that $\alpha$ particles are similarly heated. All these ions will
have significantly larger gyroradii than the protons.
In the presence of a density gradient, these ions may then excite
lower hybrid waves in the colder protons. This requires that the electron
gyroradius be less than the wavelength/2$\pi$. The waves will damp by heating electrons
in a direction parallel to the magnetic field. Unless other restrictions set in,
the waves ultimately will saturate once the electrons
are heated such that their gyroradii are no longer sufficiently small. In the region
of interest the plasma conditions are approximately magnetic field $B=1$ G and density
$n_e=10^6$ cm$^{-3}$. Then the proton and electron gyrofrequencies are
$\Omega _p=10^4$ and $\Omega_e = 1.8\times 10^7$, and the lower hybrid frequency
is $\Omega _{LH}=\sqrt{\Omega _p\Omega _e}=4\times 10^5$ (all in rad s$^{-1}$).
The electron and proton plasma frequencies are $\omega _{pe}=6\times 10^7$ and
$\omega _{pp}=1.5\times 10^6$. It will turn out that waves
are excited at low $k$ by ions moving with velocities greater than the proton thermal
speed, so the limit $\omega >> \sqrt{2}kv_i$ of the plasma dispersion function is
appropriate, and that $\alpha$ particles will be species most important for
the wave generation. This is due to their large gyroradius compared to protons
and their high abundance compared to other minor ions.

Lower hybrid waves are electrostatic ion oscillations, which can occur in
magnetic fields strong enough that the electron gyroradius is smaller
than the lower hybrid wavelength$/2\pi $. Due to this necessary magnetization of the electrons,
the waves propagate preferentially across magnetic field lines. The
parallel component of the wavevector $k_{||}/k < \omega _{pi}/\omega _{pe}$. Since
$\omega /k_{\perp} \sim \left(m_e/m_i\right) ^{1/2}\omega /k_{||}$ the wave can simultaneously be in
resonance with ions moving across the magnetic field and electrons moving along magnetic
field lines, which facilitates collisionless energy exchange between ions and electrons
on timescales much faster than that associated with Coulomb collisions.
They have also been
discussed in connection with cometary X-ray emission, electron acceleration in
solar flares, supernova remnant shock waves
and Advection Dominated Accretion Flows (ADAFS)
\citep{vaisberg83,krasno85,bingham97,shapiro99,bingham00,
begelman88,luo03,mcclements97}, and observed
in situ together with accelerated electrons at Halley's comet
\citep{gringauz86,klimov86}. A cold plasma
theory for lower-hybrid waves is given in the Appendix of \citet{laming01a}.
Here we summarize the theory with finite electron and ion temperatures.

The dispersion relation is \citep[see e.g][]{laming01b}
\begin{equation}
\omega ^2={\omega _{pe}^2\left(Ik_{||}^2/k^2\right)
/\left(1+\omega _{pe}^2/c^2k^2\right)\over 1+{\omega _{pi}^2\over
k^2v_i^2}\left(1-\phi\left(\omega\over
\sqrt{2}kv_i\right)\right)+ {\omega _{pe}^2\over
k^2v_e^2}\left(1-I\right)}
\end{equation}
where $I={m_e\over k_{\rm B}T_e}\int _0^{+\infty}
J_0^2\left(k_{\perp}v_{\perp} \over\Omega
_e\right)\exp\left(-m_ev_{\perp}^2 \over
2k_{\rm B}T\right)v_{\perp}dv_{\perp}$, $v_e^2=k_{\rm B}T_e/m_e$ and
$v_i^2=k_{\rm B}T_i/m_i$, and
$\phi\left(z\right)=-z/\sqrt{\pi}\int _{-\infty}
^{\infty}\exp\left(-t^2\right)/\left(t-z\right) dt$ is the usual plasma
dispersion function ($T_{e,i}$ and $m_{e,i}$ are electron and ion
temperatures and masses respectively and $k_{\rm B}$ is Boltzmann's constant).
Specializing to $\omega >> \sqrt{2}kv_i$ and
$\Omega _e >> k_{\perp}v_{e\perp}$, so that $I\simeq 1-k_{\perp}^2v_{e\perp}^2/
\Omega _e^2$ and $\phi \simeq 1 +i\sqrt{\pi /2}\left(\omega /kv_i\right)
\exp\left(-\omega ^2/2k^2v_i^2\right)$ and taking $\omega _{pe}>>\Omega _e$
and $c\rightarrow\infty$,
\begin{equation}
\omega =\Omega _e{k_{||}\over k}\left[1-{\Omega _e^2\over 2\omega _{pe}^2}
-{i\over 2}\sqrt{\pi\over 2}{\omega\Omega _e^2\omega _{pi}^2\over k^3v_i^3\omega _{pe}^2}
\exp\left(-\omega ^2/2k^2v_i^2\right)+ \cdots\right].
\end{equation}

The instability is illustrated schematically in Figure 1. The density
is increasing towards the right, and the density gradient is perpendicular to the
magnetic field. A local anisotropy in the ion distribution develops in the direction
formed by the cross product of density gradient and magnetic field, which in the
representation of Figure 1 is into or out of the page. This is similar to a scenario
envisaged by \citet{begelman88}, who were interested in electron-ion equilibration
in two temperature accretion flows. We consider the effect of a density gradient in
a uniform magnetic field. In such a situation the ion distribution function
is given by
\begin{equation}
f\left(v_i\right)={m_i\over 2\pi k_{\rm B}T_{i\perp}}\sqrt{m_i\over 2\pi k_{\rm B}T_{i||}}
\left(1-v_{iy}/\Omega_iL\right)
\exp\left(-{m_iv_{iz}^2\over 2k_{\rm B}T_{i||}} -{m_i\left(v_{ix}^2
+v_{iy}^2\right)\over 2k_{\rm B}T_{i\perp}}
\right),
\end{equation}
where the magnetic field $\vec{B} = B\vec{z}$, and the density gradient
$\left(dn/dx\right) \vec{x}$ is related to $L$ by $n\left(dx/dn\right) =L$.
We consider the growth of lower
hybrid waves in the protons excited by minor ions with large gyroradii.
The growth rate for each minor ion species is given by \citep{laming01a}
\begin{eqnarray}
\nonumber
&\gamma _i={Afq^2\over M}{\pi\over 2}{\omega ^3 \over nk^2}
\left(1+{\omega _{pe}^2\over \omega _{pi}^2} \cos ^2\theta\right)^{-1}
\int\vec{k}\cdot{\partial f_i\over\partial\vec{v_i}}
\delta ^3\left(\omega -\vec{k}\cdot\vec{v_i}\right)d^3\vec{v_i}\\
 &={Afq^2\over M}{\omega\over 2}\sqrt{\pi\over 2} \left(\omega \over
kv_{th\perp}\right)^3 \left(1+{\omega _{pe}^2\over \omega _{pi}^2}
\cos ^2\theta\right)^{-1}
\left[{\omega\over k\Omega _i L}-1 -{kv_{th\perp}^2\over\omega\Omega _iL}\right]
\exp\left( -{\omega ^2\over 2k^2v_{th\perp}^2}\right),
\end{eqnarray}
where $\theta$ is the angle between $\vec{B}$ and $\vec{k}$,
$v_{th\perp}=\sqrt{k_{\rm B}T_{i\perp}/m_i}$ and
$Afq^2/M$ is the product of the element abundance,
ionization fraction and charge
squared of the particular ions that excite the wave, divided by their mass in
atomic mass units. Protons are expected to have insufficiently large gyroradii
to excite lower-hybrid waves, and so the factor $Afq^2/M$ favors
$\alpha$-particles over other minor ions in the wave generation.
The term $-1$ in the square brackets represents the ion
Landau damping given by the imaginary part of the expression for $\omega$ in
equation 2. Equation 4 is similar to equation (3.20)
in \citet{begelman88}, with the identification of their drift velocity $v_{di}=
-v_{th\perp}^2/\Omega _iL$, and in their case $\omega /k << v_{th\perp}$. Their
treatment yields waves at much higher $k$ than ours does, since they consider a single
ion distribution drifting with velocity $v_{di}$ producing waves at
$k\simeq\omega/v_{di}$, whereas we treat the effect at a single point of a continuum of
different Maxwellians with differing densities (but otherwise the same) producing
waves at $k\simeq\left(\omega/v_{th\perp}\right)\left(r_g/L\right)$. This
difference is appropriate in view of the fact that they are interested in plasmas with
ion pressure $>>$ magnetic pressure $>>$ electron pressure where the plasma itself
amplifies the magnetic field, while here we are concerned
with magnetic pressure $>>$ ion and electron pressures where the magnetic field is
imposed externally. \citet{begelman88} also
considered the effects of magnetic curvature. We assume that the necessary gradients
in magnetic field are much less likely to exist in the low $\beta$ plasma of the solar
coronal hole. We find the wavevector $k_{max}$ where the growth rate is
maximized in the direction perpendicular to $\vec{B}$ (i.e. $\theta =\pi /2$),
and the maximum growth
rate itself, which is plotted in units of $\omega\times Afq^2/M$ in Figure 2.

If the waves are driven to marginal stability
$k_{max}\simeq\Omega_e/v_{e\perp}\simeq\omega r_g/Lv_{i\perp}$,
where
$v_{i\perp}$ is the perpendicular ion velocity, equation 1 can be written
\begin{equation}
v_{e\perp}^2={\Omega _e^2v_{i\perp}^2\left(1+\left(1-I\right)\omega _{pe}^2/
k^2v_{e\perp}^2\right)\left(1+\omega _{pe}^2/c^2k^2\right)\over
\omega_{pe}^2I\left(k_{||}^2/k^2\right)
+\omega _{pp}^2\left(1+\omega _{pe}^2/c^2k^2\right)}{L^2\over r_g^2}.
\end{equation}
At saturation where
$k_{\perp}v_{e\perp}/\Omega_e\simeq 1$, $I\simeq 1/2$, and putting $k_{||}/k=\cos\theta$,
\begin{equation}
v_{e\perp}^2={\Omega _e^2v_{i\perp}^2\left(1+\omega _{pe}^2/2\Omega _e^2\right)
\left(1+\omega _{pe}^2v_{e\perp}^2/\Omega _e^2c^2\right)\over
\left(\omega_{pe}^2\cos ^2\theta\right)/2
+\omega _{pp}^2\left(1+\omega _{pe}^2v_{e\perp}^2/\Omega _e^2c^2\right)}{L^2\over r_g^2}.
\end{equation}
This is a quadratic equation for $v_{e\perp}^2$ which has lowest order solution
\begin{equation}
v_{e\perp}^2 \simeq {\omega _{pe}^2v_{i\perp}^2\over 2\omega _{pp}^2}\left(
1+{\omega _{pe}^2\over 2\omega _{pp}^2}\cos ^2\theta\right)^{-1}{L^2\over r_g^2}.
\end{equation}
The kinetic growth rate varies as $\gamma\propto\left(
1+{\omega _{pe}^2\over\omega _{pp}^2}\cos ^2\theta\right)^{-1}$, \citep{laming01a}
and averaging $v_{e\perp}^2$ over this growth rate gives
$\left<v_{e\perp}^2\right>=v_{i\perp}^2\left(1-1/\sqrt{2}\right)
\omega _{pe}^2L^2/\omega _{pp}^2r_g^2$. Thus in principle electron and ion temperatures
can equilibrate (assuming that $\left(1-1/\sqrt{2}\right)L^2/r_g^2 > 1$),
in the sense that the waves do not saturate before this is achieved.

\section{Solar Wind Ionization Balance Models}
We use an adaptation of the BLASPHEMER (BLASt Propagation
in Highly EMitting EnviRonment)\footnote{The name gives away its origin in modelling
laboratory and astrophysical shock waves.} code
\citep{laming01b,laming02,laming03b,laming03c},
which follows the time dependent
ionization balance and temperatures of a Lagrangian plasma parcel as it
expands in the solar wind.
The density $n_{iq}$ of ions of element $i$ with charge $q$ is
given by
\begin{equation} {dn_{iq}\over dt} =
n_e\left(C_{ion,q-1}n_{i~q-1}-C_{ion,q}n_{iq}\right)+
n_e\left(C_{rr,q+1} +C_{dr,q+1}\right)n_{i~q+1}-
n_e\left(C_{rr,q}+ C_{dr,q}\right)n_{iq}
\end{equation}
where $C_{ion,q}, C_{rr,q}, C_{dr,q}$ are
the rates for electron impact ionization, radiative recombination and dielectronic
recombination respectively, out of the charge
state $q$. These rates are the same as those used in the recent ionization balance
calculations of \citet{mazzotta98}, using subroutines kindly supplied by
Dr P. Mazzotta (private communication 2000). The electron density $n_e$ is determined
from the condition that the plasma be electrically neutral. The ion and electron
temperatures, $T_{iq}$ and
$T_e$ are coupled by Coulomb collisions by
\begin{equation} {dT_{iq}\over dt}= -0.13n_e{\left(T_{iq}-T_e\right)\over M_{iq}T_e^{3/2}}
{q^3n_{iq}/\left(q+1\right)\over\left(\sum _{iq} n_{iq}\right)}\left(\ln\Lambda\over 37\right)
-{4\over 3}{\gamma _{iq}U_w\over n_qk_{\rm B}}
\end{equation} and
\begin{equation}
{dT_e\over dt}= {0.13n_e\over T_e^{3/2}}\sum _{iq}{\left(T_{iq}-T_e\right)\over M_{iq}}
{q^2n_{iq}/\left(q+1\right)\over\left(\sum _{iq} n_{iq}\right)}\left(\ln\Lambda\over 37\right)
-{T_e\over n_e}\left({dn_e\over dt}\right)_{ion} - {2\over 3n_ek_{\rm B}}
{dQ\over dt} +\sum _{iq}{4\over 3}{\gamma _{iq}U_w\over n_ek_{\rm B}}.
\end{equation}
Here $M_{iq}$ is the atomic mass of the ions of element $i$ and
charge $q$ in the plasma,
and $\ln\Lambda\simeq 28$ is the Coulomb logarithm. The term in $dQ/dT$
represents plasma energy losses due to ionization and radiation. Radiation losses
can be taken from \citet{summers79}, though are generally negligible in
applications to the solar wind.
The term $-\left(T_e/n_e\right)\left(dn_e/dt\right)_{ion}$ gives the reduction
in electron temperature when the electron density increases due to ionization.
Recombinations, which reduce the electron density, do not result in an increase
in the electron temperature in low density plasmas, since the energy of the
recombined electron is radiated away (in either radiative or dielectronic recombination),
rather than being shared with the other
plasma electrons as would be the case for three-body recombination in dense
plasmas.

The last terms in equations 9 and 10 represent collisionless ion-electron
energy transfer, which we estimate assuming that the population
of lower hybrid waves is approximately steady state. Then the wave growth
rate due to the gyrating ions is equal to the wave Landau damping rate due
to the electrons, and the energy transfer rate is given by $2\gamma _i U_w$
in ergs cm$^{-3}$s$^{-1}$ where $U_w=E_0^2/8\pi$ is the wave energy density in
terms of the wave peak electric field $E_0$. A threshold electric field for lower
hybrid waves between the linear and stochastic regimes has been determined by
\citet{karney78} by numerical integration
of the Hamiltonian equations of motion, and also analytically using an ion
trapping argument, to be
\begin{equation}
E_{thr} = {1\over 4}\left(\Omega _p\over\omega\right)^{1/3}{\omega\over kc}B_0,
\end{equation}
where $B_0$ is the background magnetic field and $\Omega _p$ is the
proton gyrofrequency. This field marks the onset of proton heating as well as
electron heating by the waves. We take $E_0=E_{thr}$ (probably a conservative
assumption). Writing the electron heating
rate in terms of both collisional and collisionless processes, with $T_i>>T_e$
and dropping the $q$ subscript in specializing to one ion species alone,
\begin{equation}
{dT_e\over dt}\simeq 0.13n_e{q^2/\left(q+1\right)\over M_i}{T_in_i
\over T_e^{3/2}n_H}
\left(\ln\Lambda\over 37\right) + {\gamma _i\over n_e}{B_0^2\over 64\pi}
\left(\Omega _p\over\omega\right)^{2/3}{T_i\over m_ic^2}{\omega ^2\over
k_{max}^2v_{iy}^2}
\end{equation}
where $k_{max}$ is the wavevector where the growth rate is maximized, and
$T_i=2T_{i\perp}/3=m_iv_{i\perp}^2/3=2m_iv_{iy}^2/3$ since
$v_{i\perp}^2=v_{ix}^2+v_{iy}^2=2v_{iy}^2>>v_{i||}^2$. In our
calculations we modify $\ln\Lambda $ from its usual value to account for the
collisionless energy transfer between $\alpha$-particles and electrons,
its new value being given by
\begin{equation}
\left(\ln\Lambda\over 37\right)^{\prime}=\left(\ln\Lambda\over 37\right)+
6.23\times 10^5{T_e^{3/2}B_0^3\over n_e^2}\left({\gamma _iM_i\over\omega Afq^2}
{\omega^2\over k^2v_{iy}^2}\right),
\end{equation}
the quantity $\gamma _iM_i/\omega Afq^2$ being plotted in Figure 2.
The same value is applied to electron heating by other minor ions, and the
original value of $\ln\Lambda$ is kept for proton-electron energy transfer.
In the region of the solar wind around
$1.5R_{\sun}$ heliocentric distance, where $T_e\sim 10^6$ K, $B_0\sim 1$ G,
$n_e\sim 10^6$ cm$^{-3}$ and
$\gamma ^{\prime}=\gamma _iM_i/\omega Afq^2\left(\omega /kv_{iy}\right)^2 \sim 0.1
- 1$ for $\alpha$-particles of around 5\% abundance relative to protons,
the electron heating rate is increased typically by factors of order 100 - 1000
over that due solely to Coulomb collisions with ions heavier than H.

We use the analytic model of \citet{banaszkiewicz98} for the magnetic field
strength. We take the perpendicular velocity for all ions heavier than H
from the empirical model for O VI (model B2; equation 28)
in \citet{cranmer99} in the range $1.5R_{\sun} - 3.5R_{\sun}$.
Below this we interpolate between values of 20 km s$^{-1}$ on the solar surface
and 84 km s$^{-1}$ at $1.5R_{\sun}$. The flow speed is linearly interpolated
between its initial value at the solar surface and 130 km s$^{-1}$ at
$1.7R_{\sun}$, a value determined by \citet{giordano00}. Above $1.7R_{\sun}$
the flow speed is allowed to evolve according the the action of the adiabatic
invariant in the diverging magnetic field with the perpendicular velocity
specified as above, and above $3.5R_{\sun}$ the
perpendicular velocity is also allowed to evolve in this manner, rather than
being determined by the fits in \citet{cranmer99}. The resulting
flow speed is a very good match to the empirical models B1 and B2 in
\citet{cranmer99}, as shown in Figure 3 for various values of the initial
flow speed, determined to be in the range 3 - 60 km s$^{-1}$ by various authors
\citep{patsourakos00,hassler99,wilhelm00,gabriel03}. Figure 4 shows the electron density
in the simulations for initial flow speeds between 5 and 60 km s$^{-1}$,
compared with measurements using a diagnostic line ratio in Si VIII by
\citet{doschek97} and from UVCS polarization brightness observations given
by \citet{cranmer99}. An initial flow speed of around 10-20 km s$^{-1}$ appears
to be the best match to the various observations.

Equations 8-10 are integrated following the solar wind out from an initial
position at $1.05 R_{\sun}$ out to around $7 R_{\sun}$ by which time charge
states are frozen in. Each model follows H, He, and one minor ion, C, O, Mg,
Si, or Fe in the present work, with fractional
abundances 0.83, 0.16 and 0.01 respectively
by mass. The initial electron and ion temperatures
are $9\times 10^5$ K, which also establish the initial ionization balance.
The density at this point is taken to be $10^8$ cm$^{-3}$. After each time
step in the ionization balance, the densities and electron temperature are
modified according to the adiabatic expansion of the solar wind governed
by the magnetic field geometry and the wind velocity as specified above.

Figure 5 shows the electron
temperature profile resulting from simulations with an initial flow speed of
10 km s$^{-1}$ and various values of the lower hybrid growth rate, which
depends on the density gradient assumed. For reference, the temperature
measurements of \citet{wilhelm98} using the Mg IX 750\AA\ /706\AA\ diagnostic
line ratio are also given. The error bar on the points at 1.3 and $1.6R_{\sun}$
are estimated here from the scatter in the points on Figure 8 of
\citet{wilhelm98}. These authors use the atomic physics data of \citet{keenan84}
to derive a temperature ratio, which is consistent with coronal hole
temperatures determined by other authors \citep[e.g.][]{doschek77,doschek01}.
More recent calculations in R-matrix and distorted wave approximations
summarized by \citet{landi01} give temperatures significantly higher or lower
respectively, and are likely due to inaccuracies in these atomic data.
We base our coronal hole temperatures on works such as
\citet{doschek77} and \citet{doschek01} where the ionization balance is the
principal temperature diagnostic, and on the O VI observations of
\citet{david98}, whose electron
temperature diagnostic depends on much less controversial atomic physics.
Even if the absolute temperatures measured
by \citet{wilhelm98} cannot be interpreted with confidence, as they state in
their paper, at least the maximum variation of the electron temperature with
distance from the solar surface can be constrained by their observations.
Figure 6 shows the electron temperature profiles resulting from using the
maximum collisionless energy transfer in Figure 5 and differing initial flow speeds,
with the highest temperatures resulting from the slowest initial speeds since
more time is available for the plasma electron to be heated and the rate
of temperature decrease due to adiabatic expansion is lower.

Figures 7, 8, and 9 show the evolution of the ionization balances of O, Si,
and Fe with heliocentric distance. In each case the flow starts at a density
of $10^8$ electrons cm$^{-3}$ and a temperature of $9\times 10^5$ K. The
initial flow speed is 20 km s$^{-1}$ and the electron-ion equilibration
parameter is $\gamma ^{\prime}=\gamma _iM_i/\omega Afq^2\left(\omega ^2/k^2v_{iy}^2\right) = 0.5$.
Increased ionization commences at around $1.5R_{\sun}$, in response to the
onset of ion cyclotron heating at this location, and charge states freeze
in between 2 and $2.5R_{\sun}$ at the values found in situ by Ulysses
\citep{geiss95,ko97}.

\section{Discussion}
\subsection{Charge State Distributions}
The resulting ionization balances are given in Tables 1-4 for C, O, Mg, Si
and Fe respectively. From the C and O ionization balances in Table 1,
it is clear than only for values of the parameter
$\gamma ^{\prime}=\left(\gamma _iM_i/\omega Afq^2\right)
\left(\omega /kv_{iy}\right)^2\simeq 0.5$ and initial wind speeds of 10-20
does sufficient
electron heating occur to bring the modelled charge states into agreement
with those observed. Tables 2-4 verify that these conclusions do not change when
other elements Mg, Si, and Fe are considered.
We will discuss the significance of $\gamma ^{\prime}$ below. Before
doing so we remark on the other salient features of the model. The data of
\citet{geiss95} come from a high speed stream observed closer to the ecliptic
plane than the polar observations of \citet{ko97}, and so are slightly less
highly ionized as would be expected. Even so, both sets of observations for the
parameters discussed above agree very
well with the modelled charge state fractions for C, O, Si and Fe, at least
for the two or three charge states that dominate the distributions. Mg is
still observed to be more highly charged than the models predict. However
we note that the steady state ionization balance for Mg changes dramatically
in the temperature range $9\times 10^5$ K to $1.1\times 10^6$ K
\citep{mazzotta98}, and so we
anticipate that small changes in either the model or the atomic data for Mg
would also bring this element into agreement. We
find no need to invoke different outward flow speeds for the different
elements. C, O, Mg, Si, and Fe all flow at the flow speeds determined from O VI
Doppler dimming measurements. This is in contrast with the work of \citet{ko97},
where heavier ions needed to flow out at successively lower speeds, as would
be the case in a thermal conduction driven wind. In addition to a faster
outflow, we also model higher electron temperatures, as would be expected
since less time is available before freeze-in, and so higher electron
temperatures are required to produce the necessary ionization. Even so, the
electron temperatures modelled here when extrapolated out to 0.3-1 A.U. are
still considerably lower than those measured {\it in situ} by Helios
\citep{marsch89}. The inclusion of non-zero thermal conduction (see below)
might reduce this discrepancy, but it would seem that the conclusion of
\citep{marsch89} that ``electron temperature profiles in high-speed streams
clearly indicated and even required the existence of heating sources other
than the one related to the degradation of the electron heat flux'' supports
our idea of electron heating. We also remark
that \citet{ko97} made some different choices of atomic data. In particular
their choice of O$^{6+}$ ionization rate from \citet{lennon88} is lower
than the rate adopted here from \citet{mazzotta98} by a factor of about
0.7, and elsewhere in the literature \citep[e.g.][]{moores80,shull82}
rates for this process can be greater than that of \citet{lennon88} by
nearly a factor of 2. The assessment and validation of atomic data is a huge
task, beyond the scope of the current paper \citep[see e.g.][]{savin02},
but clearly central to
further quantitative development along the lines suggested in this paper.

Further variations in the atomic rates involved may come from nonthermal
electron distributions. Throughout this paper we have taken a Maxwellian
electron distribution, tacitly assuming that the electron-electron collision
rate is sufficiently fast to maintain such conditions. We argued above that
halo electron distributions are unlikely to exist close to the Sun because
of this collisional equilibration, and numerical estimates suggest that only
at electron densities $< 10^4-10^5$ cm$^{-3}$ could a nonthermal electron
distribution produced by lower hybrid waves survive (by equating the
collisional equilibration rate with the wind expansion rate). Such densities
are only found at or beyond the radial position where ion charge states freeze
in, and so nonthermal electrons appear unlikely to produce a significant
change to our charge state results, given the other current observational
constraints. However lower hybrid waves do appear to be
a viable means for producing the electron distributions observed in the fast
solar wind.

\subsection{Small Scale Structures?}
The determination that $\gamma ^{\prime}=
\gamma _iM_i/\omega Afq^2\left(\omega /kv_{iy}\right)^2\simeq 0.5-1$
is necessary to produce the observed charge states requires the existence
of density gradients in the fast wind with scale lengths on the order of the
$\alpha$-particle gyroradius, which is about 0.1 km at 1.5 $R_{\sun}$.
Radio scintillation observations demonstrating
the existence of such size scales in the solar wind in the ecliptic plane
have rather a long history \citep[e.g.][]{coles89,armstrong90,coles91}. These
are found perpendicular to the magnetic field, with larger size scales
(typically a factor of 10 within 6 $R_{\sun}$, becoming more isotropic at
larger distances) inferred along the radial direction. \citet{coles95}
inferred values of $\delta n_e^2$ in polar regions at solar minimum to be
around 1/10 to 1/15 of that observed in equatorial regions, but due to a lack
of knowledge of $n_e$ in polar regions, were unable to say anything about the
variation of $\delta n_e/n_e$. The density measurements reviewed in this paper
indicate electron densities in polar regions a factor of 1/2 to 1/3 of those
in equatorial regions, making $\delta n_e/n_e$ in polar regions of similar
order to, but still slightly smaller than that in equatorial regions.
Absolute values of $\delta n_e/n_e$ in coronal hole regions of interest here
have been determined observationally by \citet{ofman97} to be from 0.1
to a few times 0.1. This is smaller than the value $\sim 1$ tacitly assumed
here, the consequence of which is discussed further below.

\citet{grall97} present more data on the transition from anisotropy inside
5-6 $R_{\sun}$ on scales of order 10 km to isotropy further out, concluding
that a real change in
the microstructure rather than in Alfv\'en wave turbulence takes place, again
with reference to the ecliptic plane. \citet{feldman96} review these
interplanetary scintillation
observations together with Ulysses observations to constrain the high speed
wind structure near its coronal base, and argue that the plasma is ``sufficiently
structured to relax through generation of a drift-wave instability that
results in electrostatic waves having $k$-vectors oriented perpendicular
to B'', which is precisely the motivation for the current work. \citet{grall97}
go further and show that within 6 $R_{\sun}$ large scale turbulence is isotropic
with a Kolmogorov spectrum (structure function $\propto {\rm scale}^{5/3}$),
while smaller scale turbulence shows anisotropy with higher structure functions
($\propto $ scale) than Kolmogorov turbulence would predict. The scale at
which this transition takes place which can be interpreted \citep{woo96,woo97}
as the size of the flux tube in which the wind flows, is inferred to be of order
1 km close to the sun.

In a study of electrostatic ion cyclotron wave generation by global resonant
MHD modes, \citet{markovskii01} suggests that
structures may exist with scales down to the proton inertial length
$v_A/\Omega _p = c/\omega _{pp}$, which is of the same order of magnitude as
the structures already inferred by \citet{woo96,woo97}, and in terms of
particle gyroradii about an order of magnitude larger than our present
inference. However our evaluation of the lower hybrid wave growth rate
assumed Maxwellian distributions for the minor ions. In the presence of a
radiation field interacting with the ions, \citet{hasegawa85} show that
a ``kappa'' distribution may result;
\begin{equation}
f\left(v\right)dv = {n\over\sqrt{2\pi }v_{th}}{\Gamma\left(\kappa +1\right)
\over \kappa ^{3/2}\Gamma\left(\kappa -1/2\right)}\left[1+{v^2\over 2\kappa
v_{th}^2}\right] ^{-\kappa},
\end{equation}
where $v_{th}=\sqrt{k_{\rm B}T/m}$ is the thermal velocity, $\Gamma$ is the
usual Gamma Function, and $\kappa $ is a parameter representing the deviation
from a Maxwellian distribution, which is obtained in the limit $\kappa
\rightarrow\infty $. Observations of the $\alpha$-particle distribution
function in the high speed solar wind generally give $\kappa \simeq 3-6$
\citep{collier96,chotoo98}. In Figures 10 and 11 we plot the dimensionless
lower hybrid growth rate $\gamma _iM_i/\omega Afq^2$ and the parameter
$\gamma ^{\prime}=
\gamma _iM_i/\omega Afq^2\left(\omega /kv_{iy}\right)^2$ against $L/r_g$. From Figure
11 it may be seen the for $\kappa = 3$ values of $L/r_g$ up to about 7 give
acceptable ionization and for lower $\kappa $ even higher $L/r_g$ would be
tolerable. Protons in a kappa distribution may also be able to contribute
to the wave generation, reducing the requirement on $L/r_g$ even more.
Given the relatively safe assumption that the $\kappa$ value
for $\alpha$-particles in the coronal hole is similar to or less than that
observed in high speed wind streams at about 1 A.U., we conclude that
similar density gradients to those proposed by \citet{markovskii01} would
provide sufficient lower hybrid wave generation and electron heating to
produce the observed ionization states of minor ions in the fast wind. Thus
these observations provide further support for the mechanism of ion
cyclotron wave generation throughout the extended corona proposed by
\citet{markovskii01}, that of a global resonant MHD mode driving a
cross field current in the resonant layer which then excites electrostatic
ion cyclotron waves. We have also made rather conservative assumptions
concerning the collisionless ion-electron energy transfer. Higher wave electric
fields than given by equation 11 (up to $\sim\omega B_0/kc$)
are certainly possible \citep{karney78,bingham03}, though as the wave
amplitude increases from this point, protons begin to be heated
as well as electrons (the {\it stochastic} regime). Under favorable conditions,
the ion heating rate may be similar to the electron heating rate. Thus our
estimate of the ion-electron equipartition rate may be an underestimate
by as much as an order of magnitude, if the wave electric field can be a
factor of a few higher, and the electron Landau damping rate only reduced
by a factor 0.5. Fuller discussion of this point is beyond the scope of this
paper.

\subsection{Thermal Conduction}
The approach taken in this paper of following a Lagrangian plasma element
out through the solar wind acceleration region neglects thermal
conduction. Here we argue that such an approximation is most likely justified.
The electron temperature gradient is $\sim 10^{-4}$ K cm$^{-1}$ between
1 and 2 $R_{\sun}$ and $\sim 10^{-5}$ K cm$^{-1}$ for heliocentric distances
greater than $3 R_{\sun}$. The Spitzer-Harm heat flux is then $Q\sim 10^7$
ergs cm$^{-2}$s$^{-1}$ prior to the temperature maximum (for an initial
flow speed of 5 km s$^{-1}$; it is lower for initial speeds of 10 and 20
km s$^{-1}$) directed back towards
the sun, and oppositely directed and about an order of magnitude lower
once the wind has passed the temperature maximum. The cooling rate of the
plasma occupying a length of order $R_{\sun}$ between 2 and 2 $R_{\sun}$
heliocentric distance is then $\sim 0.6$ s$^{-1}$, assuming an average electron
density in this region of $10^5$ cm$^{-3}$. This cooling rate is much
faster than that due to adiabatic expansion, $\sim 10^{-4}$ s$^{-1}$.

However in steep temperature gradients and collisionless plasma the Spitzer-Harm
thermal conductivity is not appropriate. We estimate the collisionless heat
conduction as follows:
\citet{begelman88} give the electron velocity diffusion coefficient
in the ``moderately non-linear regime'' (i.e. corresponding to the
turbulence level just before ion trapping sets in) as
\begin{equation}
D_{||}\left(v_e\right)\simeq {e^2\left<\delta E_{||}^2\right>\over m_e^2\omega }.
\end{equation}
Putting $\delta E_{||}=E_0\sqrt{m_e/2m_i}$ and using equation 11 we find
$D_{||}\simeq e^2v_i^2B_0^2/400m_em_ic^2\omega\simeq 4\times 10^{17}\left(v_i/
200 {\rm km s}^{-1}\right)^2\left(\omega /4\times 10^5 {\rm rad s}^{-1}\right)$
cm$^2$s$^{-3}$. The time taken to accelerate an electron from rest to a parallel
velocity $v_{||}$ is $t\sim v_{||}^2/D_{||}$, which evaluates to
$\sim 1$ s taking $v_{||}=1.7\times 10^9$ cm s$^{-1}$, the electron thermal
speed at $T_e=10^7$ K. The electron-electron collision time entering the
Spitzer-Harm conductivity is around 3000 s, so in lower-hybrid turbulence
the heat conduction rate should be reduced from our former estimate by a
factor 1/3000, giving a cooling rate $2\times 10^{-4}$ s$^{-1}$, which is
now comparable with that due to adiabatic expansion. Higher degrees of lower
hybrid turbulence (see above) would decrease the conductive cooling rate even
further.

That the heat conductivity should be reduced from the Spitzer-Harm value in
steep temperature gradients is well known \citep[e.g][]{bell81,salem03}.
However our estimate of heat
conduction is significantly lower than one would normally predict. Saturated
conduction is usually defined as heat conduction where electrons carry their
own thermal energy at their thermal speed. The rationale for this is that at
faster flow speeds (relative to the ions), a Buneman instability would
develop, generating Langmuir waves that would inhibit the heat flow. Applied
to the case above, the saturated heat flux would be
$\sim 3.5\times 10^5$ ergs cm$^{-2}$s$^{-1}$. In our
situation the heat conducting electrons may also excite lower-hybrid waves with
a similar threshold drift velocity. In fact we should expect the limiting
flux to be lower than this simple estimate, as indeed it is,
because in our case not only
heat conducting electrons but also ions generate the lower-hybrid turbulence.
Any electron heating mechanism invoked to explain the observed charge
state distributions in the fast solar wind must require an anomalous thermal
conductivity to avoid the deposited heat from being conducted back to the
coronal hole, which would conflict with the SUMER temperature diagnostics of
\citet{wilhelm98} and \citet{david98}. We consider that the heating and
anomalous thermal conductivity provided simultaneously by lower-hybrid waves
to be a desirable aspect of the model, in that one mechanism provides both
features.

\section{Summary and Conclusions}
In this paper we have argued that a small amount of the energy deposited
in ions between 1.5 and 2 $R_{\sun}$ should eventually find its way to the
electrons via an instability that generates lower hybrid waves. No attempt
has been made to address the ion cyclotron heating problem, other than to
show that the density gradients we require are similar to those postulated
by \citet{markovskii01} to generate electrostatic ion cyclotron waves from
a global resonant MHD mode. As such, our line of reasoning is complementary
to that in a recent paper by \citet{cranmer03}. In addressing the larger
problem of ion cyclotron heating, these authors speculate that low frequency
Alfv\'en waves can Landau damp on electrons. This parallel heating should
produce electron beaming and discrete phase-space holes which may heat ions via
stochastic processes. Thus the ion heating derives from the electron
energization (by low frequency Alfv\'en waves), rather than the electron
heating deriving from the ion heating as in the picture presented here.

While a number of quantitative issues remain unresolved (atomic data,
thermal conduction), we believe that an explanation for the observed fast
wind elemental charge states in terms of lower hybrid wave electron heating
is certainly plausible, and should be considered along with other possibilities
already discussed in the literature \citep[e.g][]{vocks03}.
More accurate numerical models would allow us to
exploit this interpretation and allow more rigorous investigation of
density inhomogeneities in the fast solar wind, with important consequences
for the generation of ion cyclotron waves throughout the extended corona.
As well as being potentially crucial to our understanding of the acceleration
of the fast solar wind, the lower hybrid wave instability is also of interest
elsewhere in astrophysics, and the Sun and solar system offer an attractive
``laboratory'' for the exploration of collisionless plasma physics processes
that are otherwise inaccessible to experiment \citep[see e.g. the recent
debate on electron-ion equilibration in ADAFs in][]{binney03,quataert03,pariev03}.

\acknowledgments
The work was supported by NASA Contract S13783G and by the NRL/ONR Solar
Magnetism and the Earth's Environment 6.1 Research Option. I am grateful to
Steven Cranmer for enlightening discussions.

\clearpage



\plotone{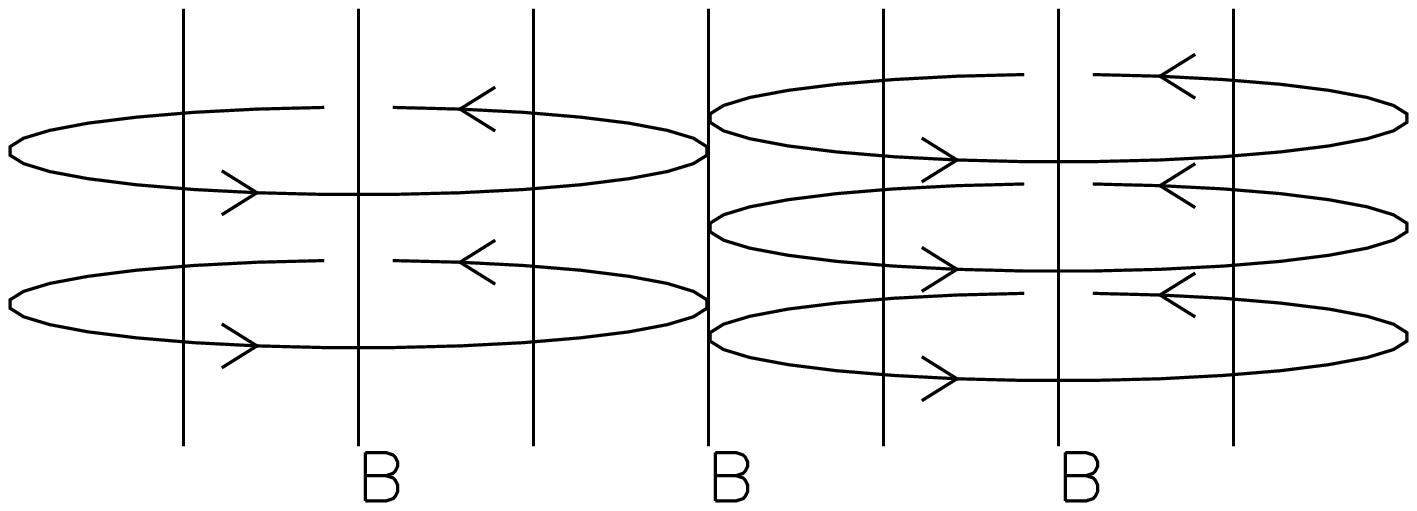}
\figcaption[f1.ps]{Schematic diagram illustrating the excitation of lower
hybrid waves in a density gradient. The magnetic field direction is vertical, and density
is increasing to the right. Gyrating ions can give a local anisotropy in the distribution
function in the direction perpendicular to both of these vector, into or out of the
page, when the characteristic length of the density gradient becomes comparable
to the ion gyroradius. \label{fig1}}

\plotone{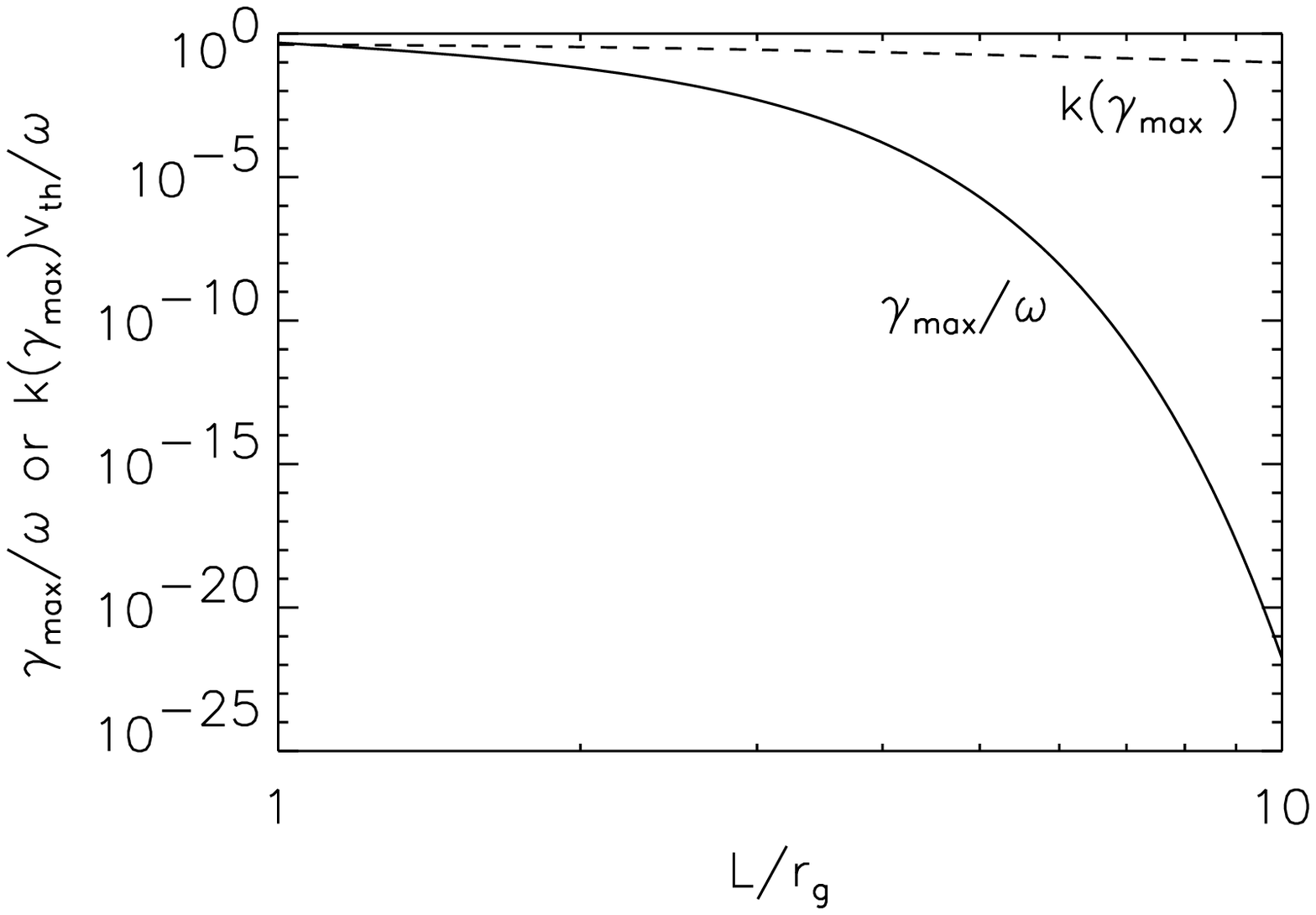}
\figcaption[f2.ps]{Plot of maximum lower-hybrid growth rate
in units of the wave frequency $\gamma _i/\omega$ (the factor $Afq^2/M$ is
omitted) and the wavevector where this maximum
is found in units of $\omega/v_{th\perp}$ against the density scale length in units of the
ion gyroradius $L/r_g$.
\label{fig2}}

\plotone{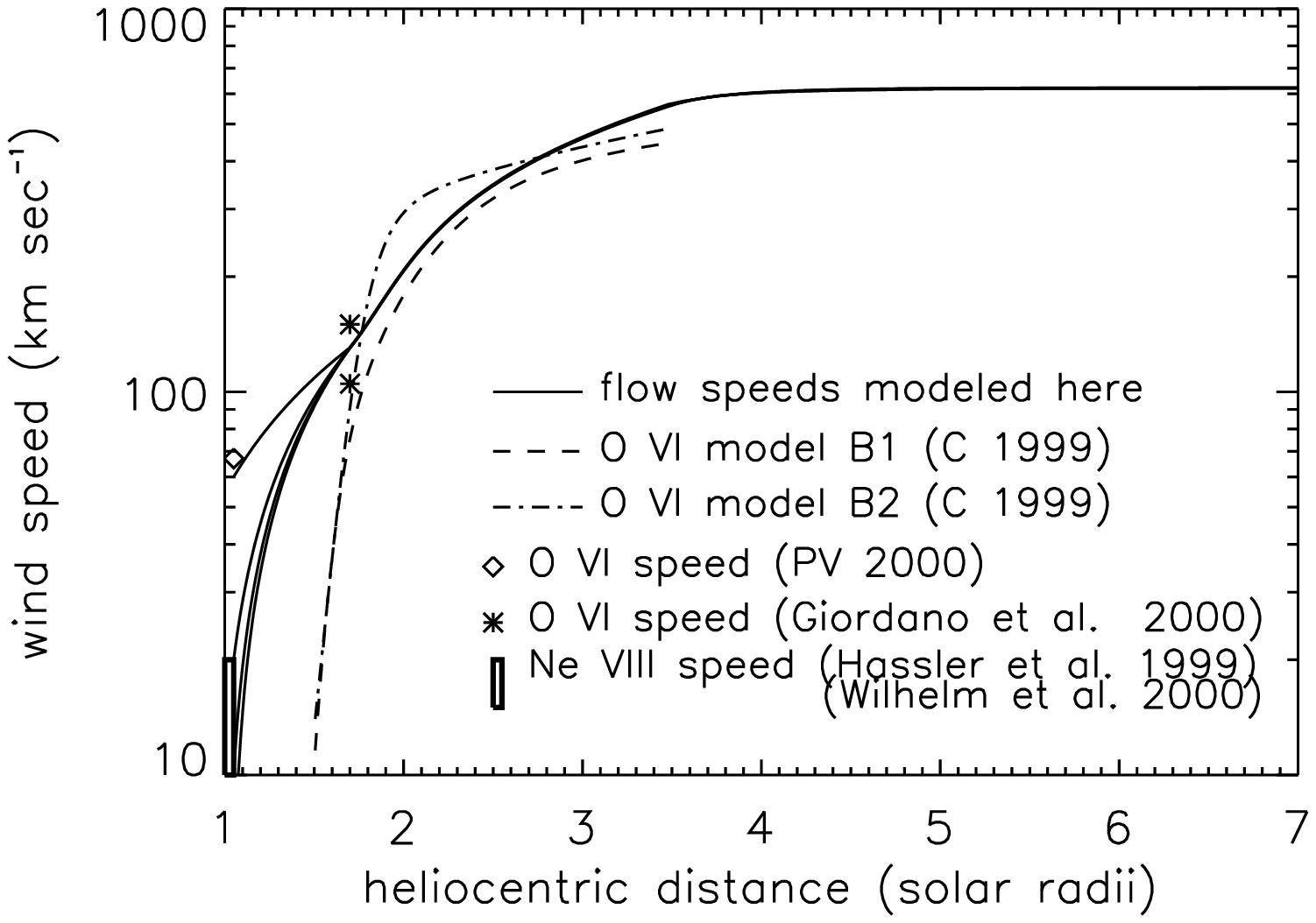}
\figcaption[f3.ps]{Plot of solar wind ion flow speeds adopted in the
ionization models. For reference we also show models B1 and B2 from
\citet{cranmer99}, some flow speeds determined from O VI Doppler dimming
observations by \citet{patsourakos00}, \citet{giordano00}, and direct Doppler
shift measurements by \citet{hassler99} and \citet{wilhelm00}.}

\plotone{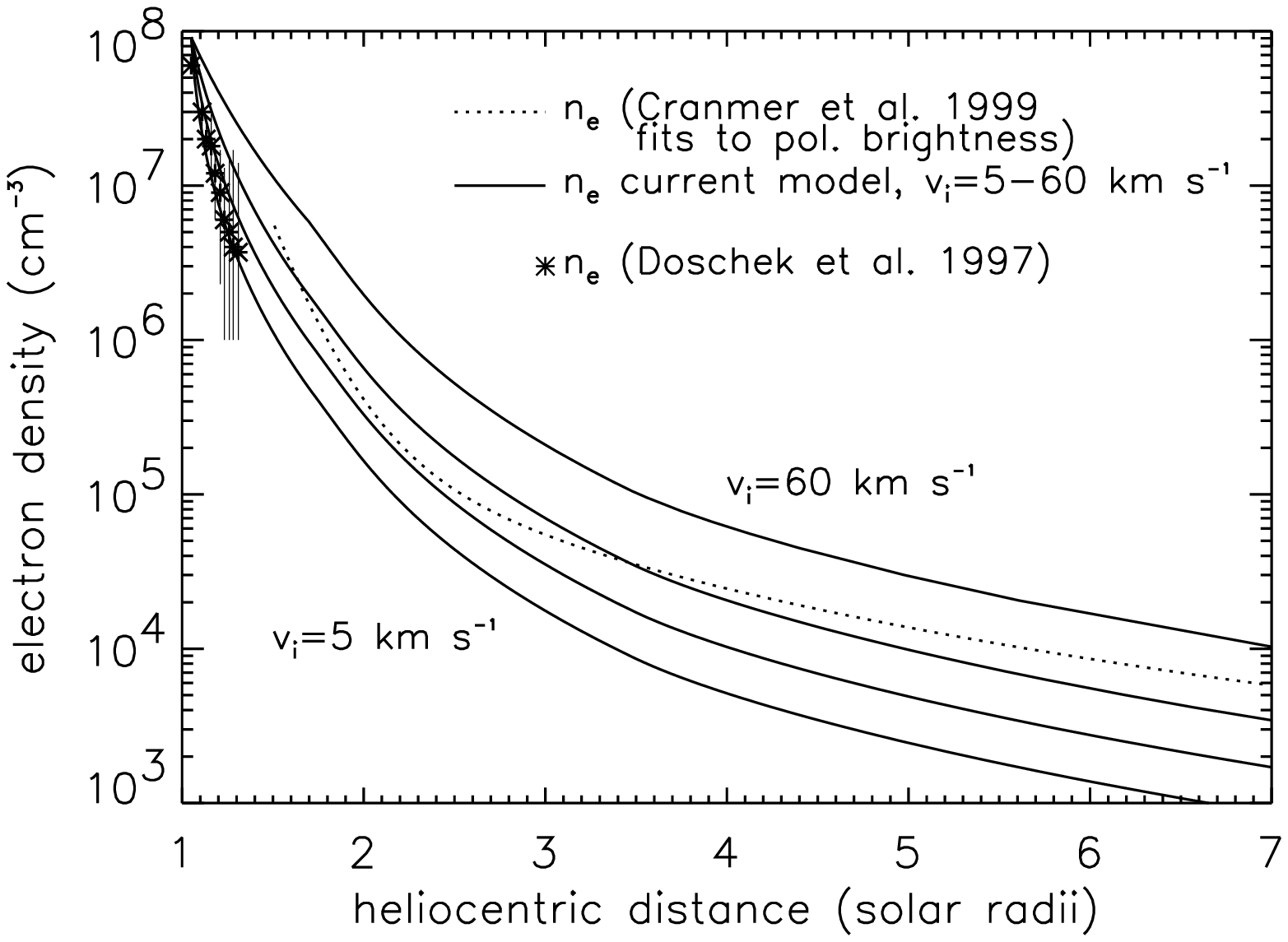}
\figcaption[f4.ps]{Plot of solar wind densities derived from the various
models with different initial flow speeds, 5, 10, 20, and 60 km s$^{-1}$,
increasing bottom to top. Densities derived from a diagnostic line ratio
in Si VIII \citep{doschek97} and polarization brightness \citep{cranmer99}
are given for comparison.}

\plotone{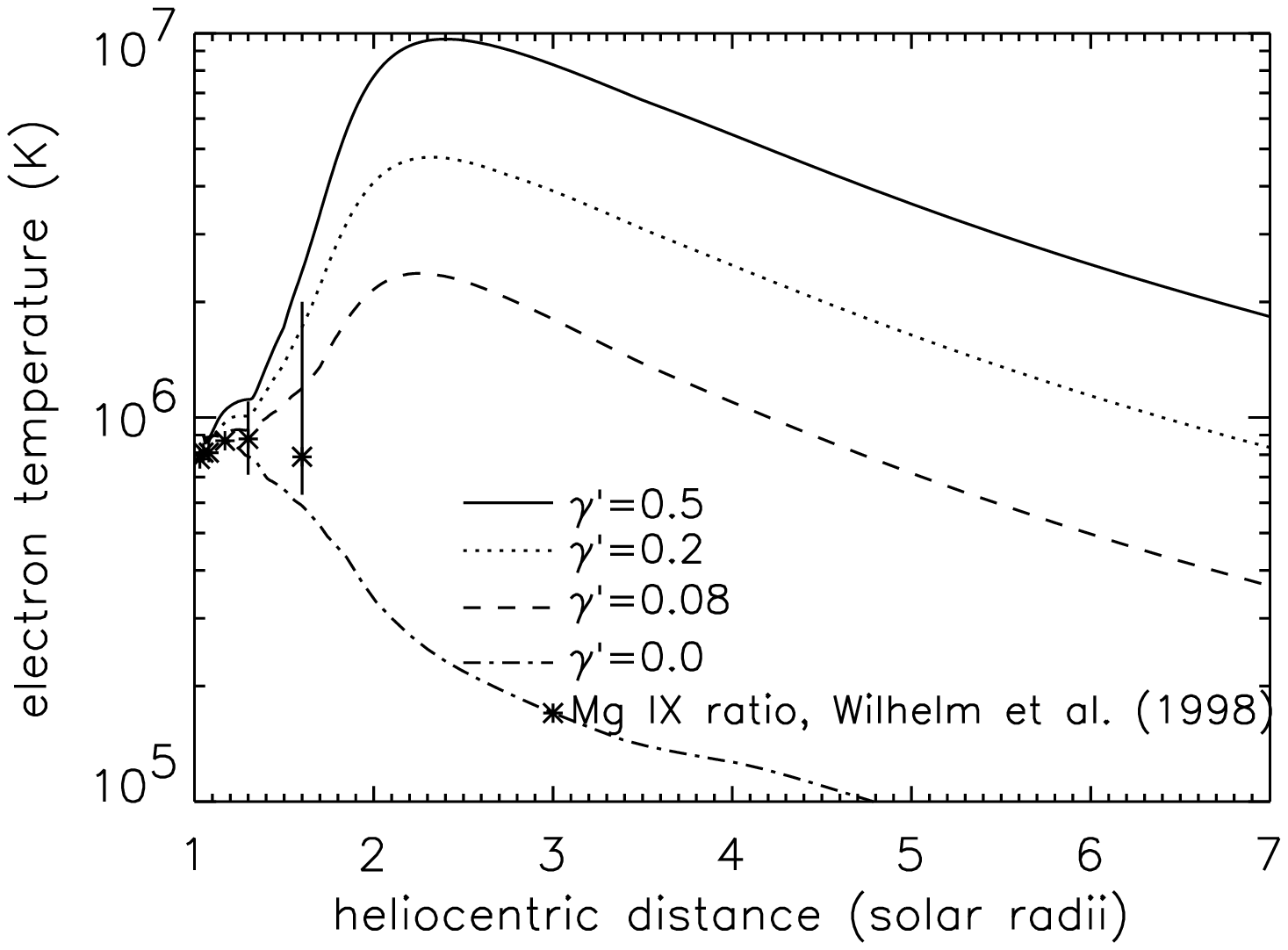}
\figcaption[f5.ps]{Plot of electron temperature variation with
heliocentric distance for models with an initial flow speed of 10 km s$^{-1}$
and varying degrees of collisionless ion-electron coupling,
$\gamma ^{\prime}=\left(\gamma _iM_i/\omega Afq^2\right)
\left(\omega /kv_{iy}\right)^2$.
Measurements from the Mg IX 706/750 temperature
sensitive ratio by \citet{wilhelm98} are given for comparison. We have estimated
error bars on their points for 1.3 and $1.6R_{\sun}$ from the scatter of points
given in their Fig. 8.}

\plotone{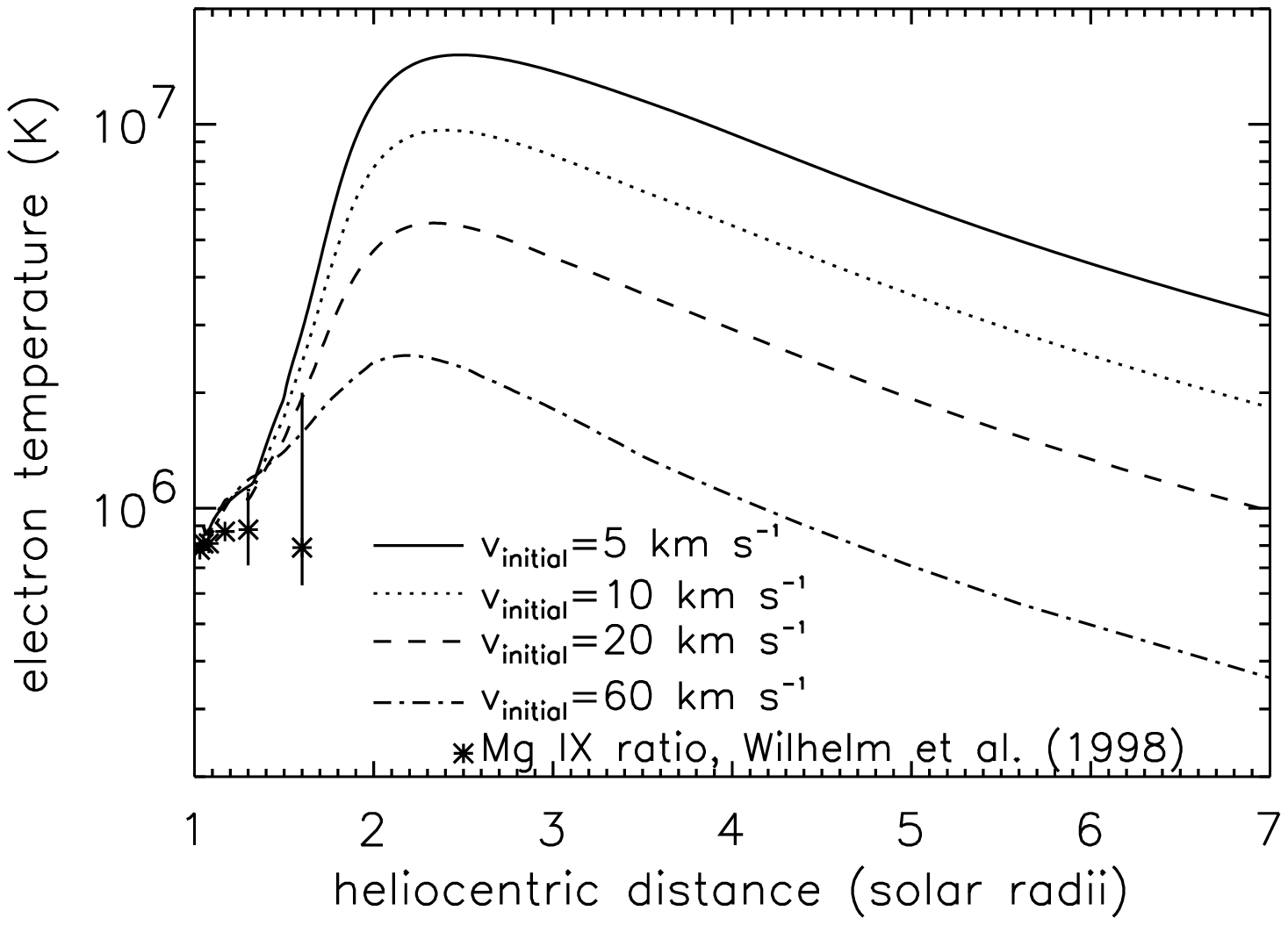}
\figcaption[f6.ps]{Plot of electron temperature variation with
heliocentric distance for models with
$\gamma ^{\prime}=\left(\gamma _iM_i/\omega Afq^2\right)
\left(\omega /kv_{iy}\right)^2 = 0.5$
and initial flow speeds varying in the range 5-60 km s$^{-1}$.
Measurements from the Mg IX 706/750 temperature
sensitive ratio by \citet{wilhelm98} are given for comparison. We have estimated
error bars on their points for 1.3 and $1.6 R_{\sun}$ from the scatter of points
given in their Fig. 8.}

\plotone{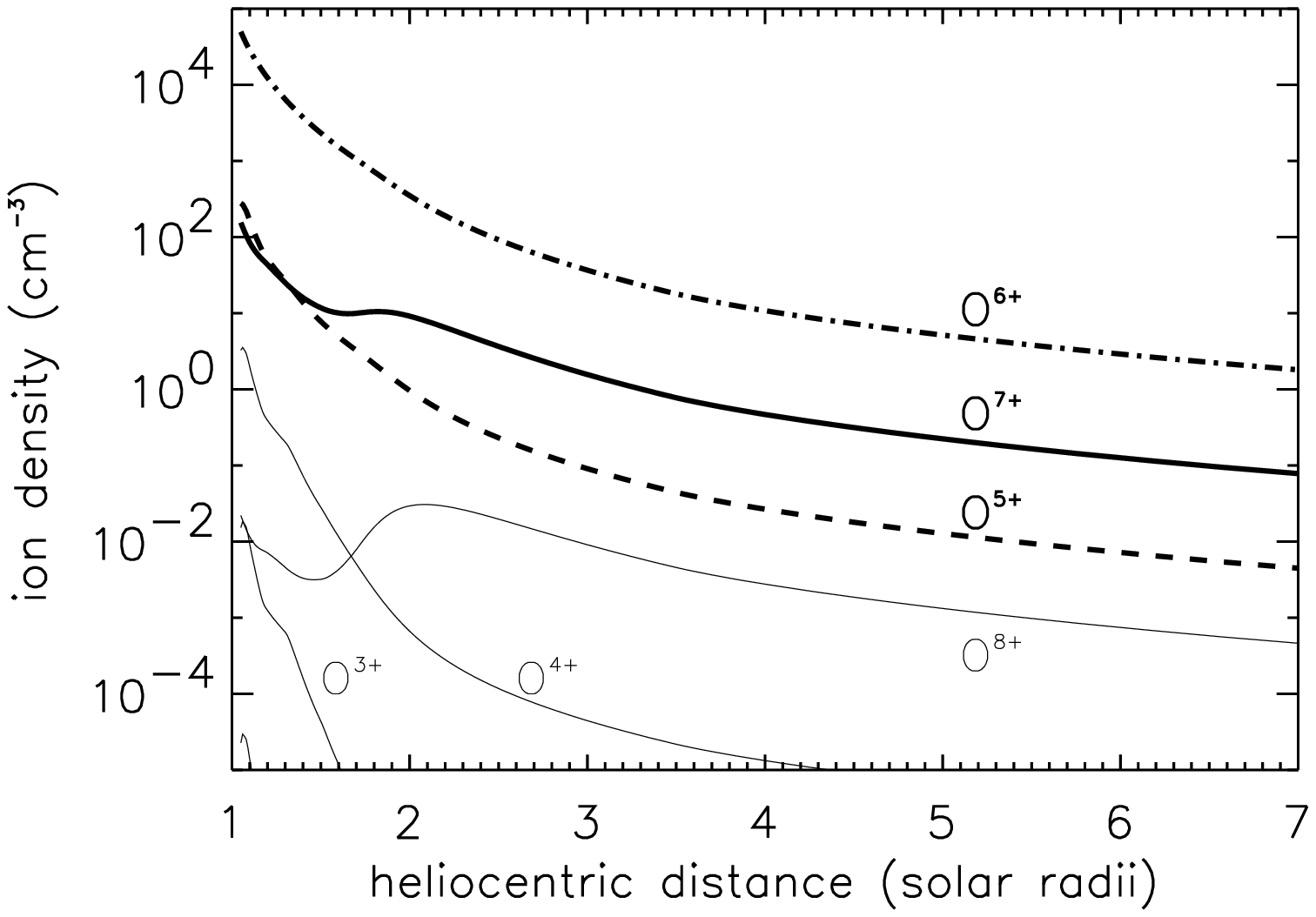}
\figcaption[f7.ps]{Plot of the evolution of the O ionization balance with
heliocentric distance for initial flow speed 20 km s$^{-1}$ and
$\left(\gamma _iM_i/\omega Afq^2\right)\left(\omega ^2/k^2v_{iy}^2\right)=0.5$,
corresponding to $L/r_g\simeq 2$. The initial ionization
balance corresponds to the coronal hole electron temperature of
$9\times 10^5$K. Increased ionization starts at about $1.5R_{\sun}$, as
ion-electron energy transfer increases in response to the strong ion cyclotron
heating at this location. Charge states are frozen in beyond a distance of
2-2.5 $R_{\sun}$, and correspond to those measured in situ by Ulysses.}
\clearpage
\plotone{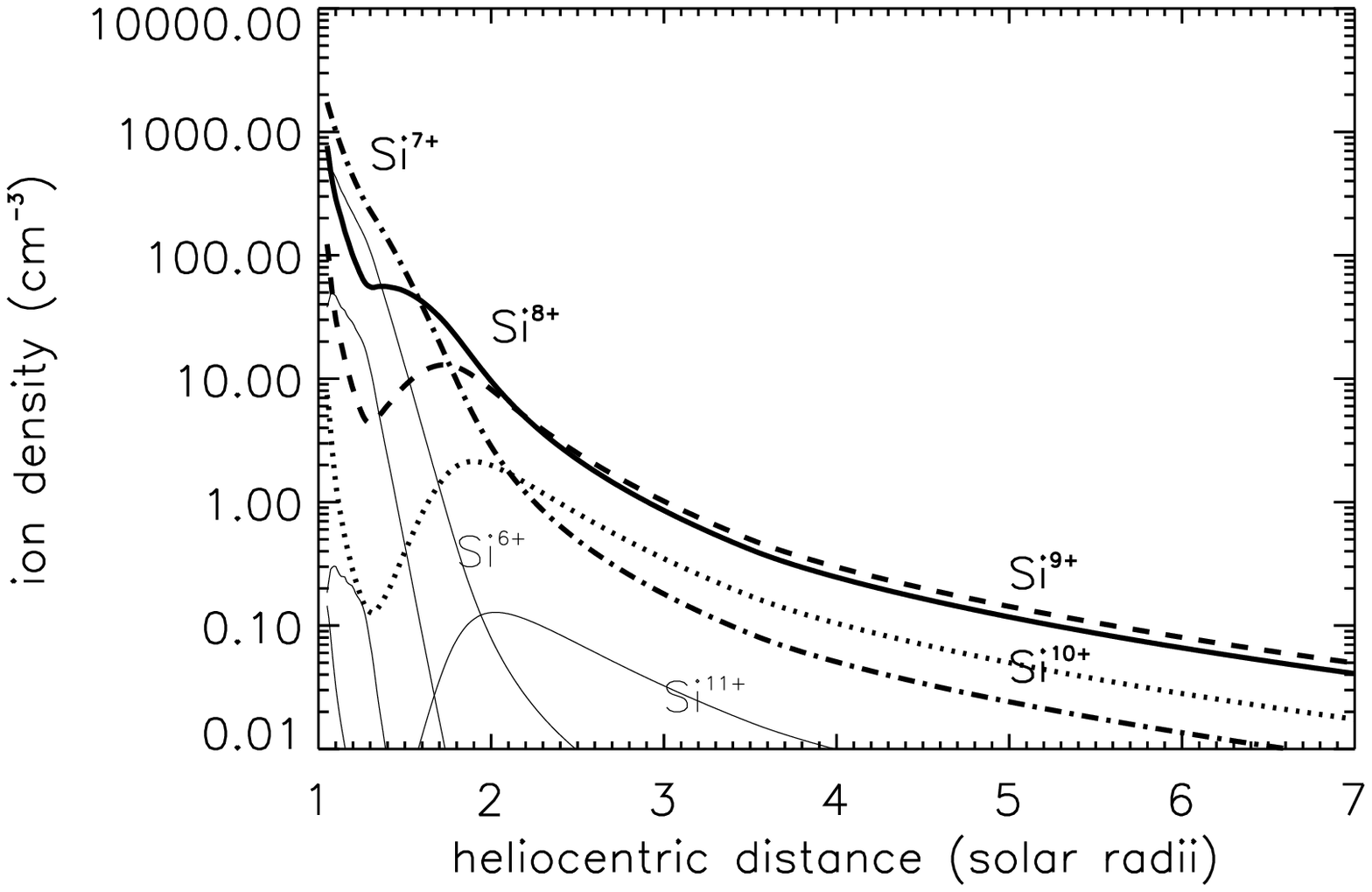}
\figcaption[f8.ps]{Plot of the evolution of the Si ionization balance with
heliocentric distance for initial flow speed 20 km s$^{-1}$ and
$\left(\gamma _iM_i/\omega Afq^2\right)\left(\omega ^2/k^2v_{iy}^2\right)=0.5$,
corresponding to $L/r_g\simeq 2$. The initial ionization
balance corresponds to the coronal hole electron temperature of
$9\times 10^5$K. Increased ionization starts at about $1.5R_{\sun}$, as
ion-electron energy transfer increases in response to the strong ion cyclotron
heating at this location. Charge states are frozen in beyond a distance of
2-2.5 $R_{\sun}$, and correspond to those measured in situ by Ulysses.}

\plotone{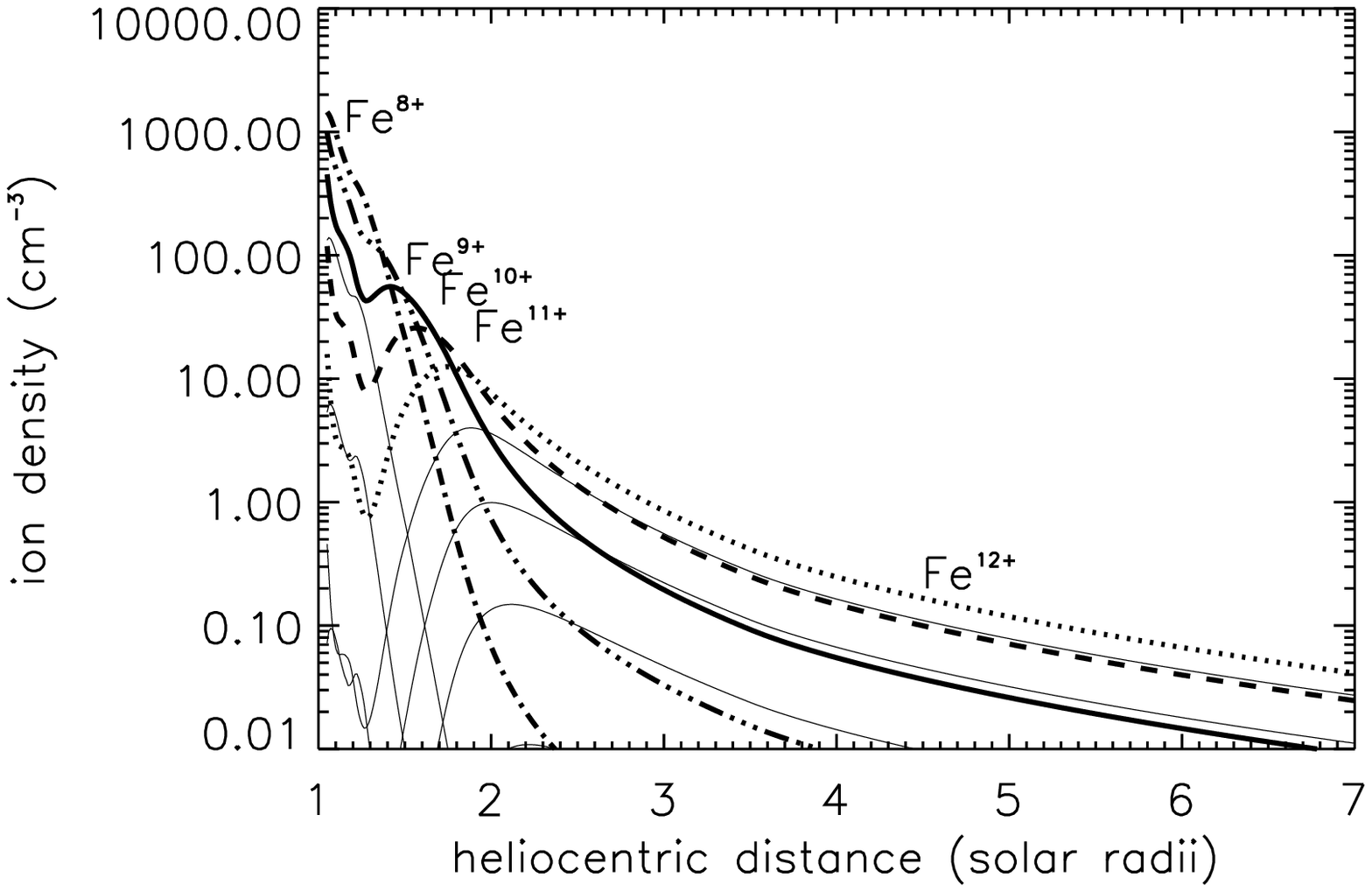}
\figcaption[f9.ps]{Plot of the evolution of the Fe ionization balance with
heliocentric distance for initial flow speed 20 km s$^{-1}$ and
$\left(\gamma _iM_i/\omega Afq^2\right)\left(\omega ^2/k^2v_{iy}^2\right)=0.5$,
corresponding to $L/r_g\simeq 2$.
The initial ionization
balance corresponds to the coronal hole electron temperature of
$9\times 10^5$K. Increased ionization starts at about $1.5R_{\sun}$, as
ion-electron energy transfer increases in response to the strong ion cyclotron
heating at this location. Charge states are frozen in beyond a distance of
2-2.5 $R_{\sun}$, and correspond to those measured in situ by Ulysses.}

\plotone{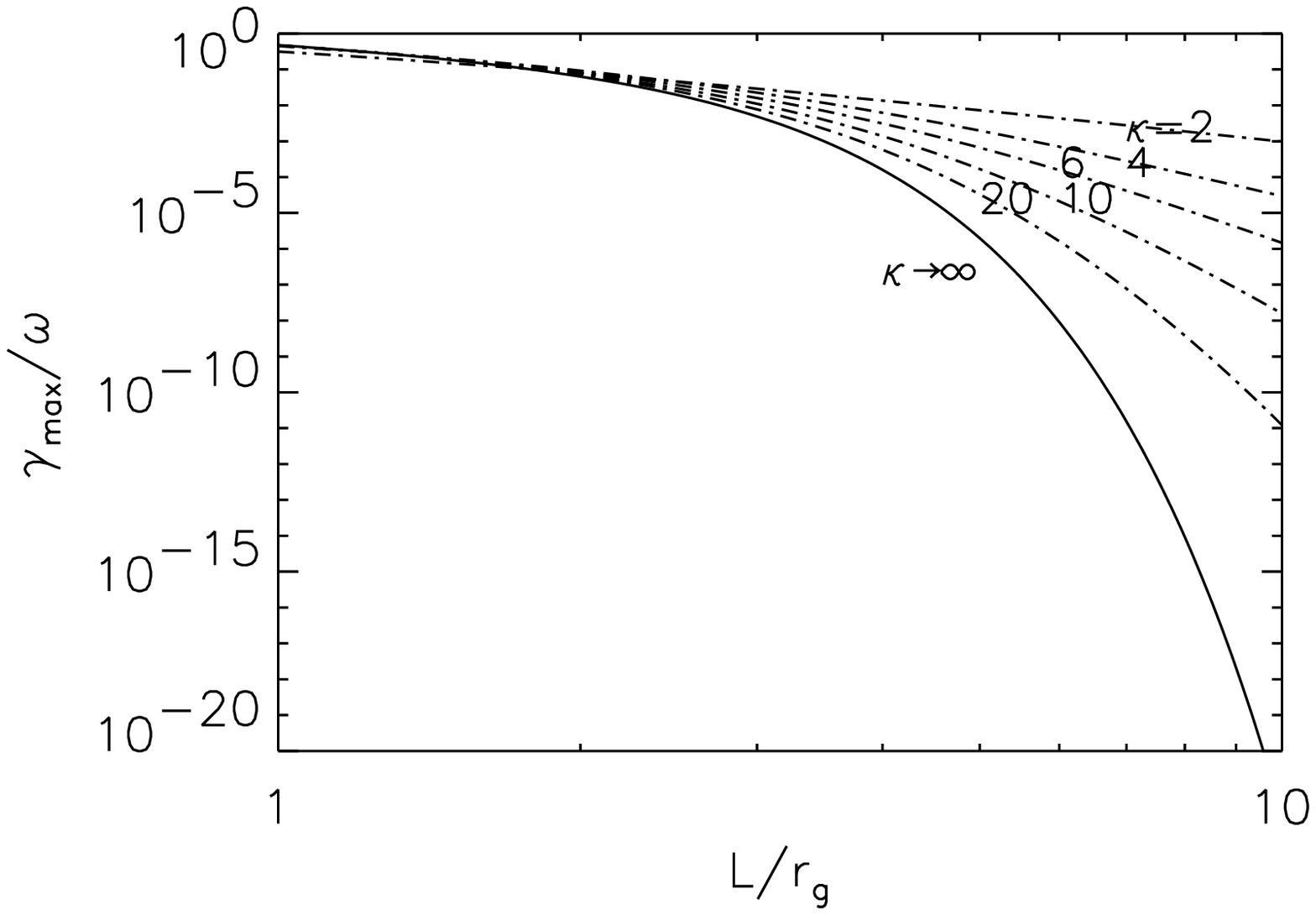}
\figcaption[f10.ps]{Plot of maximum lower-hybrid growth rate
in units of the wave frequency $\gamma _iM_i/\omega Afq^2$
against the density scale length in units of the
ion gyroradius $L/r_g$, for Maxwellian $\kappa\rightarrow\infty$ and $\kappa
=2$, 4, 6, 10, and 20.}

\plotone{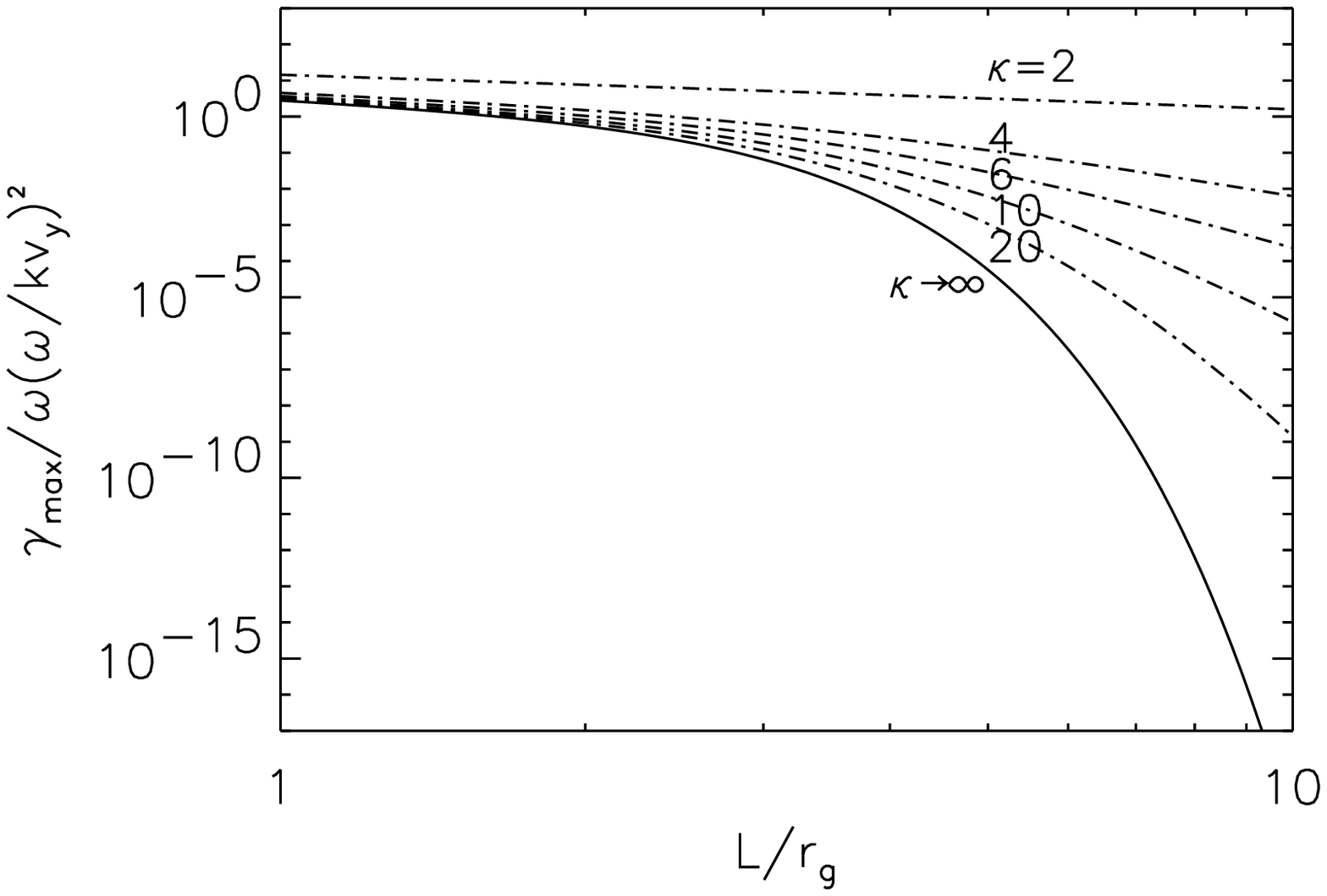}
\figcaption[f11.ps]{Plot of $\gamma ^{\prime}=
\left(\gamma _iM_i/\omega Afq^2\right)\left(\omega /kv_{iy}\right)^2$
in units of the wave frequency against the density scale length in units of the
ion gyroradius $L/r_g$, for Maxwellian $\kappa\rightarrow\infty$ and $\kappa
=2$, 4, 6, 10, and 20.}

\clearpage

\begin{deluxetable}{crrrr}
\tabletypesize{\scriptsize}
\tablecaption{C and O Ionization Fractions. \label{tableco}}
\tablewidth{0pt}
\tablehead{
\colhead{$v_{start}$} & \colhead{$\gamma ^{\prime}$}
& \colhead{C$^{4+}$/C$^{5+}$} & \colhead{C$^{5+}$/C$^{6+}$}
& \colhead{O$^{6+}$/O$^{7+}$}\\
\colhead{km s$^{-1}$} & &}
\startdata
2.5 & 0.0 & 0.86& 9.0& 375\\
2.5 & 0.048& 0.82& 8.4& 100\\
2.5 & 0.2 & 0.78& 7.7& 50\\
2.5 & 0.5 & 0.74& 7.2& 35\\
2.5 & 1.0 & 0.64 & 6.4& 24\\
\\
5& 0.0 & 0.84& 8.9& 361\\
5& 0.08 & 0.81& 8.4& 148\\
5& 0.2& 0.71& 7.3& 46\\
5& 0.5 & 0.63& 6.3& 24\\
5& 1.0 & 0.57 & 5.7& 18\\
\\
10 & 0.0 & 0.80& 8.7& 340\\
10 & 0.08& 0.80& 8.6& 282\\
10 & 0.2& 0.68& 7.3& 65\\
10 & 0.5 & 0.53& 5.5& 20\\
10 & 1.0 & 0.45& 4.5& 12\\
\\
20 & 0.0 & 0.76& 8.4& 317\\
20 & 0.08 & 0.77& 8.5& 334\\
20 & 0.2&  0.69& 7.7& 141\\
20 & 0.5 & 0.48& 5.5& 25\\
20 & 1.0 & 0.35& 3.7& 11\\
\\
60 & 0.0 & 0.68& 7.9& 259\\
60 & 0.08 & 0.67& 7.8& 257\\
60 & 0.2& 0.64& 7.6& 222\\
60 & 0.5 & 0.51& 6.2& 73\\
60 & 1.0& 0.31& 3.8& 28\\
\\
\citet{geiss95} & & 0.42& 5.0& 30\\
\citet{ko97}& & 0.48& 5.3& 32\\
\enddata
\end{deluxetable}

\begin{deluxetable}{crrrrr}
\tabletypesize{\scriptsize}
\tablecaption{Mg Ionization Fractions ($\gamma ^{\prime}=0.5$). \label{tablemg}}
\tablewidth{0pt}
\tablehead{
\colhead{$v_{start}$} & \colhead{Mg$^{6+}$} &
\colhead{Mg$^{7+}$} & \colhead{Mg$^{8+}$} &\colhead{Mg$^{9+}$}
&\colhead{Mg$^{10+}$}\\
\colhead{km s$^{-1}$}}
\startdata
2.5 &0.053& 0.31 & 0.45 & 0.14& 0.043 \\
5& 0.036 & 0.26& 0.47& 0.18& 0.052\\
10 & 0.012 & 0.16& 0.49& 0.27& 0.073\\
20 & 0.006 & 0.11& 0.47& 0.32& 0.090\\
60 & 0.006 & 0.10& 0.48& 0.32& 0.088\\
\\
\citet{ko97} & 0.028& 0.12& 0.23& 0.23& 0.40\\
\enddata
\end{deluxetable}

\begin{deluxetable}{crrrrr}
\tabletypesize{\scriptsize}
\tablecaption{Si Ionization Fractions ($\gamma ^{\prime}=0.5$). \label{tablesi}}
\tablewidth{0pt}
\tablehead{
\colhead{$v_{start}$} & \colhead{Si$^{7+}$} &
\colhead{Si$^{8+}$} & \colhead{Si$^{9+}$} &\colhead{Si$^{10+}$}
&\colhead{Si$^{11+}$}\\
\colhead{km s$^{-1}$}}
\startdata
2.5 &0.36& 0.43 & 0.15 & 0.019& 0.001 \\
5& 0.24 & 0.45& 0.24& 0.045& 0.003\\
10 & 0.12 & 0.40& 0.37& 0.10& 0.009\\
20 & 0.068 & 0.34& 0.43& 0.15& 0.015\\
60 & 0.095 & 0.38& 0.40& 0.11& 0.007\\
\\
\citet{geiss95} & 0.08& 0.31& 0.41& 0.19& 0.01\\
\citet{ko97} & 0.056& 0.21& 0.43& 0.23& 0.054\\
\enddata


\end{deluxetable}

\begin{deluxetable}{crrrrrr}
\tabletypesize{\scriptsize}
\tablecaption{Fe Ionization Fractions ($\gamma ^{\prime}=0.5$). \label{tablefe}}
\tablewidth{0pt}
\tablehead{
\colhead{$v_{start}$} &
\colhead{Fe$^{9+}$} & \colhead{Fe$^{10+}$} &\colhead{Fe$^{11+}$}
&\colhead{Fe$^{12+}$} & \colhead{Fe$^{13+}$} &\colhead{Fe$^{14+}$}\\
\colhead{km s$^{-1}$}}
\startdata
2.5 & 0.30& 0.34& 0.19& 0.06& 0.009& 0.001\\
5 & 0.14& 0.30& 0.30& 0.18& 0.049& 0.009\\

10 & 0.034 & 0.15& 0.28& 0.31& 0.16& 0.050\\
20 & 0.012 & 0.077& 0.21& 0.35& 0.23& 0.095\\
60 & 0.023 & 0.11& 0.26& 0.37& 0.18& 0.051\\
\\
\citet{geiss95} & 0.04& 0.17& 0.28& 0.26& 0.16& 0.05\\
\citet{ko97} & & 0.16& 0.25& 0.28& 0.16\\
\enddata


\end{deluxetable}


\begin{thebibliography}{}
\bibitem[Armstrong et al.(1990)]{armstrong90}Armstrong, J. W., Coles, W. A.,
Kojima, M., \& Rickett, B. J. 1990, \apj, 358, 685
\bibitem[Banaszkiewicz, Axford \& McKenzie(1998)]{banaszkiewicz98} Banaszkiewicz, M.,
Axford, W. I., \& McKenzie, J. F. 1998, \aap, 337, 940
\bibitem[Begelman \& Chiueh(1988)]{begelman88}Begelman, M. C., \& Chiueh, T.
1988, \apj, 332, 872
\bibitem[Bell, Evans \& Nicholas(1981)]{bell81}Bell, A. R., Evans, R. G.,
\& Nicholas, D. J. 1981, \prl, 46, 243
\bibitem[Bingham et al.(1997)]{bingham97}Bingham, R., Dawson, J. M., Shapiro, V. D.,
Mendis, D. A., \& Kellett, B. J. 1997, Science, 275, 49
\bibitem[Bingham et al.(2000)]{bingham00}Bingham, R., Kellett, B. J., Dawson, J. M.,
Shapiro, V. D., \& Mendis, D. A. 2000, ApJS, 127, 233
\bibitem[Bingham et al.(2003)]{bingham03}Bingham, R., Kellett, B. J., Cairns,
R. A., Tonge, J., \& Mendon\c ca, J. T. 2003, \apj, 595, 279
\bibitem[Binney(2003)]{binney03}Binney, J. 2003, \mnras,  submitted,
astro-ph/0308171
\bibitem[Chen, Esser, \& Hu(2003)]{chen03}Chen, Y., Esser, R., \& Hu, Y. 2003,
\apj, 582, 467
\bibitem[Chotoo et al.(1998)]{chotoo98}Chotoo, K., Collier, M. R., Galvin, A.
B., Hamilton, D. C., \& Gloeckler, G. 1998, \jgr, 103, 17441
\bibitem[Coles \& Harmon(1989)]{coles89}Coles, W. A., \& Harmon, J. K. 1989.
\apj, 337, 1023
\bibitem[Coles et al.(1991)]{coles91}Coles, W. A., Lui, W., Harmon, J. K.,
\& Martin, C. L. 1991, JGR, 96, 1745
\bibitem[Coles et al.(1995)]{coles95}Coles, W. A., Grall, R. R., Klinglesmith,
M. T., \& Bourgois, G. 1995, JGR, 100, 17069
\bibitem[Collier et al.(1996)]{collier96}Collier, M. R., Hamilton, D. C.,
Gloeckler, G., Bochsler, P., \& Sheldon, R. B. 1996, \grl, 23, 1191
\bibitem[Cranmer \& van Ballegooijen(2003)]{cranmer03}Cranmer, S. R., \&
van Ballegooijen, A. A. 2003, \apj, 594, 573
\bibitem[Cranmer(2000)]{cranmer00}Cranmer, S. R. 2000, \apj, 532, 1197
\bibitem[Cranmer et al.(1999)]{cranmer99}Cranmer, S. R. et al. 1999, \apj,
511, 481,
\bibitem[Cranmer, Field, \& Kohl(1999)]{cranmer99a}Cranmer, S. R., Field,
G. B., \& Kohl, J. L. 1999, \apj, 518, 937
\bibitem[David et al.(1998)]{david98}David, C., Gabriel, A. H., Bely-Dubau, F.,
Fludra, A., Lemaire, P., \& Wilhelm, K. 1998, \aap, 336, L90
\bibitem[Doschek \& Feldman(1977)]{doschek77}Doschek, G. A., \& Feldman, U. 1977,
\apj, 212, L143
\bibitem[Doschek et al.(1997)]{doschek97}Doschek, G. A., Warren, H. P., Laming,
J. M., Mariska, J. T., Wilhelm, K., Lemaire, P., Sch\"uhle, U., \& Moran, T. G.
1997, \apj, 482, L109
\bibitem[Doschek et al.(2001)]{doschek01}Doschek, G. A., Feldman, U., Laming,
J. M., Sch\"uhle, U., \& Wilhelm, K. 2001, \apj, 546, 559
\bibitem[Esser \& Edgar(2000)]{esser00}Esser, R., \& Edgar, R. J. 2000, \apj, 532, L71,
\bibitem[Esser \& Edgar(2001)]{esser01}Esser, R., \& Edgar, R. J. 2001, \apj, 563, 1062,
\bibitem[Feldman et al.(1975)]{feldman75}Feldman, W. C., Asbridge, J. R., Bame, S. J.,
Montgomery, M. D., \& Gary, S. P. 1975, JGR, 80, 4181
\bibitem[Feldman et al.(1996)]{feldman96}Feldman, W. C., Barraclough, B. L.,
Phillips, J. L., \& Wang, Y.-M. 1996, \aap, 316, 355
\bibitem[Gabriel, Bely-Dubau, \& Lemaire(2003)]{gabriel03}Gabriel, A. H.,
Bely-Dubau, F., \& Lemaire, P. 2003, \apj, 589, 623
\bibitem[Geiss et al.(1995)]{geiss95}Geiss, J., et al. 1995, Science, 268, 1033
\bibitem[Giordano et al.(2000)]{giordano00}Giordano, S., Antonucci, E., Noci, G.,
Romoli, M., \& Kohl, J. L. 2000, \apj, 531, L79
\bibitem[Grall et al.(1997)]{grall97}Grall, R. R., Coles, W. A., Spangler, S. R.,
Sakurai, T., \& Harmon, J. K. 1997, JGR, 102, 263
\bibitem[Gringauz et al.(1986)]{gringauz86} Gringauz, K. I., et al. 1986, \nat,
321, 282
\bibitem[Hasegawa, Mima \& Duong-van(1985)]{hasegawa85}Hasegawa, A., Mima, K.,
\& Duong-van, M. 1985, \prl, 54, 2608
\bibitem[Hassler et al.(1999)]{hassler99}Hassler, D. M., Dammasch, I. E., Lemaire, P.,
Brekke, P., Curdt, W., Mason, H. E., Vial, J.-C., \& Wilhelm, K. 1999, Science, 283,
810
\bibitem[Karney(1978)]{karney78}Karney, C. F. F. 1978., Phys. Fluids, 21, 1584
\bibitem[Keenan et al.(1984)]{keenan84}Keenan, F. P., Kingston, A. E., Dufton,
P. L., Doyle, J. G., \& Widing, K. G. 1984, Solar Physics, 94, 91
\bibitem[Keenan (1984)]{keenan84a}Keenan, F. P. 1984, Solar Physics, 91, 27
\bibitem[Klimov et al.(1986)]{klimov86}Klimov, S. S., et al. 1986, \nat, 321,
292
\bibitem[Ko et al.(1997)]{ko97}Ko, Y.-K., Fisk, L. A., Geiss, J., Gloeckler, G.,
\& Guhathakurta, M. 1997, Solar Physics, 171, 345
\bibitem[Kranosel'skikh et al.(1985)]{krasno85}Krasnosel'skikh, V. V., Kruchina,
E. N., Thejappa, G., \& Volokitin, A. S. 1985, \aap, 149, 323
\bibitem[Laming(2001a)]{laming01a} Laming, J. M. 2001a, \apj, 546, 1149
\bibitem[Laming(2001b)]{laming01b} Laming, J. M. 2001b, \apj, 563, 828
\bibitem[Laming \& Feldman(2003)]{laming03a} Laming, J. M., \& Feldman, U. 2003,
\apj, 591, 1257
\bibitem[Laming \& Grun(2002)]{laming02}Laming, J. M., \& Grun, J. 2002,
\prl, 89, 125002
\bibitem[Laming \& Grun(2003)]{laming03b}Laming, J. M., \& Grun, J. 2003,
Physics of Plasmas, 10, 1614
\bibitem[Laming \& Hwang(2003)]{laming03c}Laming, J. M., \& Hwang, U. 2003,
\apj, 597, 347
\bibitem[Landi et al.(2001)]{landi01}Landi, E., Doron, R., Feldman, U., \&
Doschek, G. A. 2001, \apj, 556, 912
\bibitem[Lennon et al.(1988)]{lennon88}Lennon, M, Bell, K. L., Gilbody, H. B., Hughes, J.
G., Kingston, A. E., Murray, M. J., \& Smith, F. J. 1988, J. Phys. Chem. Ref. Data, 17,
1285
\bibitem[Li(2003)]{li03}Li, X. 2003, \aap, 406, 345
\bibitem[Luo, Wei, \& Feng(2003)]{luo03}Luo, Q. Y., Wei, F. S., \& Feng, X. S. 2003,
\apj, 584, 497
\bibitem[Markovskii(2001)]{markovskii01}Markovskii, S. A. 2001, \apj, 557, 337
\bibitem[Marsch et al.(1989)]{marsch89}Marsch, E., Pilipp, W. G., Thieme, K. M.,
\& Rosenbauer, H. 1989, JGR, 94, 6893
\bibitem[Marsch(1991)]{marsch91}Marsch, E. 1991, in Physics of the Inner Heliosphere,
ed. R. Schwenn \& E. Marsch, (Berlin: Springer), 45
\bibitem[Mazzotta et al.(1998)]{mazzotta98}Mazzotta, P., Mazzitelli, G., Colafranceso, S.,
\& Vittorio, N. 1998, \aaps, 133, 403
\bibitem[McClements et al.(1997)]{mcclements97} McClements, K. G., Dendy, R.
O., Bingham, R., Kirk, J. G., \& Drury, L. O'C. 1997, \mnras, 291, 241
\bibitem[Moores, Golden, \& Sampson(1980)]{moores80}Moores, D. L., Golden, L. B., \&
Sampson, D. H. 1980, J. Phys. B. 13, 385
\bibitem[Neugebauer et al.(1996)]{neugebauer96}Neugebauer, M., Goldstein, B. E.,
Smith, E. J., \& Feldman, W. C. 1996, JGR, 101, 17047
\bibitem[Ofman et al.(1997)]{ofman97}Ofman, L., Romoli, M., Poletto, G., Noci, G.,
\& Kohl, J. L. 1997, \apj, 491, L111
\bibitem[Ofman \& Davila(2001)]{ofman01}Ofman, L., \& Davila, J. M. 2001, \apj, 553, 935
\bibitem[Pariev \& Blackman(2003)]{pariev03}Pariev, V. I., \& Blackman, E. G.
2003, \mnras, submitted, astro-ph/0310167
\bibitem[Patsourakos \& Vial(2000)]{patsourakos00}Patsourakos, S., \&
Vial, J.-C. 2000, \aap, 359, L1
\bibitem[Quataert(2003)]{quataert03}Quataert, E. 2003, \mnras, submitted,
astro-ph/0308451
\bibitem[Reisenfeld et al.(2001)]{reisenfeld01}Reisenfeld, D. B., Gary, S. P., Gosling,
J. T., Steinberg, J. T., McComas, D. J., Goldstein, B. E., \& Neugebauer, M. 2001,
JGR, 106, 5693
\bibitem[Salem et al.(2003)]{salem03}Salem, C., Hubert, D., Lacombe, C., Bale, S.
D., Mangeny, A., Larson, D. E., \& Lin, R. P. 2003, \apj, 585, 1147
\bibitem[Savin \& Laming(2002)]{savin02}Savin, D. W., \& Laming, J.M. 2002,
\apj, 566, 1166
\bibitem[Schwartz, Feldman, \& Gary(1981)]{schwartz81}Schwartz, S. J., Feldman,
W. C., \& Gary, S. P. 1981, JGR, 86, 4574
\bibitem[Shapiro et al.(1999)]{shapiro99}Shapiro, V. D., Bingham, R., Dawson, J .M.,
Dobe, Z., Kellett, B. J., \& Mendis, D. A. 1999, JGR, 104, 2537
\bibitem[Shull \& van Steenberg(1982)]{shull82}Shull, J. M., \& van Steenberg, M. 1982,
\apjs, 48, 95
\bibitem[Summers \& McWhirter(1979)]{summers79}Summers, H. P., \& McWhirter, R. W. P.,
J. Phys. B., 12, 2387
\bibitem[Vaisberg et al.(1983)]{vaisberg83} Vaisberg, D. L., Galeev, A. A.,
Zastenker, G. N., Klimov, S. I., Nozdrachev, M. N., Sagdeev, R. Z., Sokolov, A.
Y., \& Shapiro, V. D. 1983, Sov. Phys. JETP, 58, 716
\bibitem[Vocks \& Mann(2003)]{vocks03}Vocks, C., \& Mann, G. 2003, \apj, 593,
1134
\bibitem[von Steiger et al.(2000)]{vonsteiger00} von Steiger, R., et al. 2000, JGR,
105, 27217
\bibitem[Wilhelm et al.(1998)]{wilhelm98}Wilhelm, K., Marsch, E., Dwivedi, B.,
Hassler, D. M., Lemaire, P., Gabriel, A. H., \& Huber, M. C. E. 1998, \apj,
500, 1023
\bibitem[Wilhelm et al.(2000)]{wilhelm00}Wilhelm, K., Dammasch, I. E., Marsch, E., \&
Hassler, D. M. 2000, \aap, 353, 749
\bibitem[Woo(1996)]{woo96} Woo, R., 1996, \nat, 379, 321
\bibitem[Woo \& Habbal(1997)]{woo97} Woo, R., \& Habbal, S. R. 1997, \apj, 474,
39
\end{thebibliography}
\end{document}